\documentclass[pre,twocolumn,showpacs,floatfix]{revtex4}
\usepackage{graphicx}
\usepackage{amsmath}
\usepackage{amssymb}
\usepackage{subfigure}
\usepackage{multirow}
\newcommand{\ww}[1]{\underline{\underline{{\bf #1}}}}

\newcommand{\be}{\begin{equation}}
\newcommand{\ee}{\end{equation}}
\newcommand{\bea}{\begin{eqnarray}}
\newcommand{\eea}{\end{eqnarray}}
\newcommand{\ba}{\begin{aligned}}
\newcommand{\ea}{\end{aligned}}
\newcommand{\bma}{\begin{bmatrix}}
\newcommand{\ema}{\end{bmatrix}}

\newcommand{\bi}{\begin{itemize}}
\newcommand{\ei}{\end{itemize}}

\newcommand{\ave}[1]{\langle #1 \rangle}
\newcommand{\mustat}{\mu^{\text{stat}}}
\newcommand{\phistat}{\Phi^{\text{stat}}}
\newcommand{\mustatinf}{\mu_{\infty}^{\text{stat}}}
\newcommand{\phistatinf}{\Phi_{\infty}^{\text{stat}}}
\newcommand{\phisoinf}{\Phi_{\infty}^{\text{iso}}}
\newcommand{\phiso}{\Phi^{\text{iso}}}
\newcommand{\ka}{\kappa_1}
\newcommand{\kb}{\kappa_2}
\renewcommand{\imath}{i}
\begin{document}

\title{Frictionless bead packs have macroscopic friction, but no dilatancy.}

\author{Pierre-Emmanuel Peyneau}
\email{pierre-emmanuel.peyneau@lcpc.fr}
\author{Jean-No\"el Roux}
\affiliation{Universit\'e Paris-Est, LMSGC\footnote{
Laboratoire des Mat\'eriaux et des Structures du G\'enie Civil is a joint laboratory depending on Laboratoire Central des Ponts et Chauss\'ees, \'Ecole Nationale des Ponts et Chauss\'ees and Centre National de la Recherche Scientifique.}, Institut Navier, 2 all\'ee Kepler, Cit\'e Descartes, 77420 Champs-sur-Marne, France}

\date{\today}

\begin{abstract}
The statement of the title is shown by numerical simulation of homogeneously sheared assemblies of frictionless, nearly rigid beads in the quasistatic limit. Results coincide for steady flows at constant shear rate $\dot\gamma$  in the limit of small $\dot\gamma$ and static approaches, in which packings are equilibrated under growing deviator stresses. The internal friction angle $\varphi$, equal to $5.76\pm 0.22$ degrees in simple shear, is independent of average pressure $P$ in the rigid limit and stems from the ability of stable frictionless contact networks to form stress-induced anisotropic fabrics. No enduring strain localization is observed. Dissipation at the macroscopic level results from repeated network rearrangements, like the effective friction of a frictionless slider on a bumpy surface. Solid fraction $\Phi$ remains equal to the random close packing value $\simeq 0.64$ in slowly or statically sheared systems. Fluctuations of stresses and volume are observed to regress in the large system limit. Defining the inertial number as  $I=\dot\gamma\sqrt{m/(aP)}$, with $m$ the grain mass and $a$ its diameter, both internal friction coefficient $\mu^*=\tan\varphi$ and volume $1/\Phi$ increase as powers of $I$ in the quasistatic limit of vanishing $I$, in which all mechanical properties are determined by contact network geometry. The microstructure of the sheared material is characterized with a suitable parametrization of the fabric tensor and measurements of coordination numbers.
\end{abstract}


\maketitle

\section{Introduction}
Packings of particles appear in a variety of fields of condensed matter physics and material science, such as granular materials~\cite{HHL98,HW04,GRMH05}, powders~\cite{Ca05}, or concentrated, non-colloidal suspensions~\cite{SP05,OBR06}. Such systems are macroscopically disordered, and share many common features in their rheological behavior.
One is a certain shear stress threshold, above which they roughly qualify as a fluid, and below which they might be regarded as solid.
In assemblies of particles with purely repulsive force laws, interactions often do not introduce any stress scale, and the threshold only
involves some \emph{ratio} of stress components, whence a behavior often expressed as a \emph{friction law}.
Another basic property shared by many particulate systems is the existence of a specific value of the particle density,
above which the material cannot flow. The viscosity of a dense suspension diverges as the solid fraction
$\Phi$ approaches some value $\Phi^*$, often regarded~\cite{Brady93} as identical to the random close packing one,
$\Phi_{\text{RCP}}$ ($\Phi_{\text{RCP}} \simeq 0.64$ for identical spherical balls~\cite{CC87}).
Shearing and volumetric strains are coupled in granular media, which, once densely packed, cannot deform without expanding:
this is the dilatancy property, first introduced by Reynolds~\cite{RE85}. Once the shear strain reaches a large enough value,
granular packs can continuously deform, like perfectly plastic materials, under constant stresses while keeping a constant solid fraction $\Phi_c$:
this state of steady plastic flow does not depend on initial conditions and is known in soil mechanics as the \emph{critical state}~\cite{DMWood}.
Friction and dilatancy are coupled in granular materials by the stress-dilatancy relations, as proposed, \emph{e.g.} by Rowe~\cite{DMWood,PWR62}.

It is tempting to try and identify simple, model systems apt to explore the microscopic origin of those broadly defined rheological features.
To this end, discrete particle numerical simulation, for granular materials~\cite{CS79,HHL98,RoCh05}, or suspensions~\cite{BrBo88},
has now become a widespread research tool. Thus friction laws in model granular assemblies in steady shear flows, with some inertial effects,
were simulated~\cite{Dacruz05,Hatano07}, and stress-dilatancy relations were tested~\cite{TER06}.
Many results were obtained on sphere packings~\cite{TH00,SUFL04}, which, long investigated in order to characterize their geometry~\cite{CC87,BH93},
are now studied with complete mechanical models. Thus it has been checked~\cite{OSLN03,DTS05,iviso1}
that the random close packing state of monosized spheres is apparently uniquely defined if enduring agitation inducing traces of
crystalline order is avoided in the assembling stage. The macroscopic (or internal) friction coefficients,
and their relation to micromechanical parameters, including intergranular friction, have been evaluated from numerical simulations~\cite{TH00,SUFL04}.

Despite recent advances, some open gaps and unanswered questions can be pointed out in the literature.
The accurate and detailed characterization of frictionless systems under isotropic loads~\cite{OSLN03,DTS05},
in which static equilibrium states are studied, and few parameters are introduced, contrast with the more general investigations
of the behavior of granular systems with intergranular friction~\cite{TH00,SUFL04,TER06}, which are most often carried out
by dynamical methods involving inertia effects, and involve quite a few additional parameters. In those latter studies,
the limit of frictionless grains is not really treated with the desirable accuracy. Yet, frictionless packings, albeit reported to exhibit
singular properties~\cite{CR2000,RC02,TW00}, incorporate basic geometric effects that are common to suspensions and dry granular systems,
even though they are supplemented by viscous flow effects in the first case, by intergranular friction and inertia in the second case.

In order to clarify issues that have not been settled, the present paper is devoted to a numerical study of
\emph{frictionless bead packings}, subjected to homogeneous shear, and addresses the following questions.
Can frictionless packs sustain shear stresses in static equilibrium states as well as in slow,
steady flow, and do static and dynamic friction coefficients coincide~? Do fluctuations on measured stresses or strain rates regress in the large system limit~?
What can be said about characteristic densities
$\Phi_{\text{RCP}}$, $\Phi^*$, $\Phi_c$~? How do classical approaches of dilatancy~\cite{RE85,GODI98}, and the way it couples to friction~\cite{TER06},
apply in such a simple case~?

The paper is organized in four main parts. Section~\ref{sec:model} describes the model material and the numerical simulation setup,
specifying the boundary condition and initial states used in static and dynamic approaches. Section~\ref{sec:global} reports on the main results about the
macroscopic behavior -- macroscopic friction and dilatancy -- and their dependence on the dimensionless control parameters identified in Section~\ref{sec:model}.
Section~\ref{sec:micro} investigates the packing microstructure and the force networks, in connection with macroscopic mechanical properties,
with, in particular, a detailed characterization of anisotropy. Section~\ref{sec:dis} is a discussion.

\section{Model material and numerical experiments\label{sec:model}}
\subsection{System and interactions}\label{sec:system}
We consider packings of equal-sized spherical beads of diameter $a$ and mass $m$, enclosed in a cuboidal simulation cell.

Beads interact in their contacts where only normal forces $F_N$ are transmitted,
which are modeled as a sum of an elastic term and a viscous one, as in many numerical
studies of granular systems (see e.g., Refs.~\cite{TH00,SEGHL02,iviso1,XO06}).
The elastic force $F_N^e$ is related to the
normal deflection $h$ of the contact by the Hertz law~\cite{JO85},
\be
F_N^e=\frac{\tilde E}{3}\sqrt{a}h^{3/2},
\label{eq:Hertz}
\ee
where $\tilde E$ is a notation for $E/(1-\nu^2)$, $E$ is the Young modulus of the solid material the
spherical grains are made of, and $\nu$ its Poisson ratio.
Eq.~\eqref{eq:Hertz} attributes to contacts a variable spring constant $K_N = \textrm{d}F_N^e/\textrm{d}h = \tilde E \sqrt{a}h^{1/2}/2$.

The viscous normal force opposes the relative normal velocity $\delta V_N=\dot h$ of contacting beads, and is chosen as
\be
F_N^v= \zeta \sqrt{2mK_N}\delta V_N =\zeta (m\tilde E)^{1/2}(ah)^{1/4}\delta V_N,
\label{eq:visc}
\ee
with a constant coefficient $\zeta$. The same form of the viscous force was used in~\cite{SRSvHvS05,iviso1}. Although~\eqref{eq:visc} is devoid of a physical justification, some kind of dissipation is required (a granular material is not a conservative system), and consequently, the
influence of $\zeta$ on the simulation results has to be carefully assessed. One attractive feature of the force law~\eqref{eq:Hertz} and~\eqref{eq:visc} is the resulting velocity-independent coefficient of restitution $e_N$ in binary collisions. Most simulations reported here were done with $\zeta$ values such that $e_N$ is close to zero.

Particle rotations play no role and are ignored, as frictionless spherical objects behave like point particles interacting with central forces.

The equations of motion for the particles, given by Newton's law, as in all molecular dynamics (MD) methods, are to be numerically solved with standard time discretization schemes~\cite{AT87}. The time step used in the computations is a small fraction of the characteristic period of oscillations for the stiffest contact.

\subsection{Boundary conditions, stress and strain control\label{sec:bc}}
We use different simulation procedures in which some strain, or strain rate, and stress components are externally imposed to the system.

In order to avoid wall effects and to determine easily the intrinsic constitutive laws that apply in the large system limit,
the simulation cell has periodic boundary conditions. The edges of the cell have
lengths $(L_\alpha)_{1 \leq \alpha \leq 3}$ along the three orthogonal axes of coordinates. Unlike the cell, the material may undergo some shear strain, imposed with the Lees-Edwards procedure~\cite{AT87}. Adding this possibility to the potential shrinking
deformations along the three axes of coordinates, four independent strain components are considered in the different simulation steps and
methods we are using in this work. The procedures defined below consist in choosing to fix some of them to zero or to a constant
value while prescribing the values of stresses
conjugate to the others. Table~\ref{tab:sim} recapitulates those choices for the three different simulation procedures.
\subsubsection{Initial assembling process: procedure O\label{sec:OMD}}
In a preliminary step, the system is first prepared by isotropic compression of a loose ``gas'' of particles. The corresponding
procedure, denoted as ``O'' (like ``origin''), is the one applied in~\cite{iviso1} to prepare isotropic packings.
Global shear strain $\gamma$ is kept equal to zero, while the system shrinks along all three directions, until a mechanical equilibrium state is
reached for which all three diagonal components $\sigma_{\alpha\alpha}$ of the Cauchy stress tensor~\cite{BR90,CMNN81} are equal to a set pressure value $P$.
The system is deemed equilibrated when all forces compensate to zero, with a tolerance set to $10^{-4}Pa^2$ on each particle, and
each diagonal stress component is equal to $P$ with a relative error smaller than $10^{-4}$, while
the kinetic energy per particle does not exceed $10^{-8}Pa^3$. Those isotropic equilibrated configurations are the ``random close packing states'', as studied
in~\cite{OSLN03,DTS05,iviso1}.
\subsubsection{Controlled shear rate: procedure D\label{sec:DMD}}
Initial configurations produced with method O may then be subjected to a simple shear deformation, in which a macroscopic
motion along direction 1 is set up, with velocity gradient, on average, along direction 2 (by the Lees-Edwards procedure), while $L_3$ and $L_1$ are fixed.
$L_2$ is allowed to fluctuate in order to maintain $\sigma_{22}$ equal to $P$ on average (with very small fluctuations). The macroscopic shear rate is denoted as $\dot{\gamma}$. This defines procedure
``D'' (for dynamically sheared). It was implemented in a very similar way in~\cite{RoRoWoNaCh06}. One then records the time-averaged shear stress $\tau = \ave{\sigma_{12}}$, as well as the sample volume.
It is important to note that Lees-Edwards boundary conditions are fully compatible
with either a linear velocity profile or very heterogeneous strain fields. With this procedure, shear strain $\gamma$ is set equal to the ratio of the offset along axis $x_1$ of the neighbor copy of the simulation cell in the $x_2$ direction, to the length $L_2$. Consequently, due to fluctuations in $L_2$, the time derivative of $\gamma$ is not strictly equal to $\dot{\gamma}$ at all times.
\subsubsection{Static approach, controlled shear stress: procedure S\label{sec:SMD}}
In the limit of small $\dot\gamma$, results of procedure D simulations
should be comparable to static computations, in which the system equilibrates under
an externally imposed shear stress. To compare static and dynamic measurements (possible differences between ``static'' and ``dynamic''
friction coefficients or threshold shear stresses in similar systems are mentioned in~\cite{Hatano07}, and discussed in~\cite{Dacruz02,XO06}),
we also implemented a completely stress-controlled, quasistatic procedure, denoted as
``S'' (for static). In procedure S, increasing values of shear stress $\tau$ are stepwise applied, by increments $\delta \tau = 0.005\times P$,
to the initially isotropic configurations obtained with procedure O, while the prescribed value of all three diagonal components
$\sigma_{\alpha\alpha}$  is the initial pressure $P$. $\dot \gamma$, unlike in procedure D, is not constant. It satisfies a dynamical equation designed to impose a prescribed value $\tau$ to $\sigma_{12}$.
For each value of $\tau$, one waits until a satisfactory equilibrium state is reached (with the same tolerance levels as
in procedure O). The calculation is stopped when the packing does not equilibrate with the current value of $\tau$ after $5\times 10^7$ MD time steps.
The largest $\tau$ value for which an equilibrium state was obtained is kept as an estimate of the shear stress threshold for onset of flow.

\begin{table}
\begin{ruledtabular}
\begin{tabular}{cccc}
  Stress/strain control & Procedure O & Procedure D & Procedure S \\
  \hline
  $\sigma_{11}\,/\,L_{1}$ & $\sigma_{11} = P$ & constant $L_{1}$ & $\sigma_{11} = P$ \\
  $\sigma_{22}\,/\,L_{2}$ & $\sigma_{22} = P$ & $\sigma_{22} = P$ & $\sigma_{22} = P$ \\
  $\sigma_{33}\,/\,L_{3}$ & $\sigma_{33} = P$ & constant $L_3$ & $\sigma_{33} = P$ \\
  $\sigma_{12}\,/\,\dot\gamma$ & $\dot\gamma= 0$ & constant~$\dot{\gamma}$ & $\sigma_{12} = \tau$ \\
\end{tabular}
\caption{Choice of imposed stresses or strain rates in the three simulation procedures O, D, and S.
\label{tab:sim}}
\end{ruledtabular}
\end{table}
\subsection{Dimensional analysis, state parameters and geometric limit\label{sec:dimanalysis}}
Assuming homogeneous steady states are observed in large enough samples, with shear rate $\dot \gamma$ and
normal stress $P$, then, by dimensional analysis~\cite{RoCh05,Dacruz05,iviso1} all dimensionless state variables, such
as solid fraction $\Phi$ or average stress ratio $\ave{\sigma_{12}/\sigma_{22}}$ only depend on three dimensionless parameters.

The first one, the \emph{inertial number}, $I~=~\dot\gamma\sqrt{\frac{m}{aP}}$, characterizes the importance of inertial effects in dense granular flows~\cite{Gdr04,Dacruz05,Hatano07} and plays a central role for these systems~\cite{JFP06,iviso2}. The quasistatic limit is the limit of $I\to 0$, which we will systematically explore.

The importance of contact deformation is characterized by the second dimensionless parameter, a \emph{stiffness number}
which we define, like in~\cite{iviso1}, as $\kappa = (E/P)^{2/3}$. $\kappa$ is such  that the typical contact deflection $h$ under pressure $P$ is proportional $\kappa^{-1}a$ with a prefactor close to 1~\cite{iviso1}.
In order to enable comparison of macroscopic elastic properties with experimental results, we set $E = 70$~GPa and $\nu = 0.3$ (glass elastic constants). The pressure levels chosen, $P=10$~kPa and $P=100$~kPa, then respectively correspond to $\kappa=\ka\simeq 3.9\times 10^4$ and
 $\kappa=\kb\simeq 8.4\times 10^3$. These two values of $\kappa$ were reported to be large enough for the limit of
rigid grains, \emph{i.e.}, of $\kappa\to+\infty$, to be approached with good accuracy in the case of static packings~\cite{iviso1}.

Finally, the third dimensionless parameter is the level of \emph{viscous damping} $\zeta$, which appears in a viscous force and should not play a major role in the quasistatic limit.

Table~\ref{tab:dimless} sums up the values of dimensionless control parameters used in the present numerical study.
\begin{table}
\begin{ruledtabular}
\begin{tabular}{ccc}
 $I$ & $\kappa$ & $\zeta$ \\ \hline
 $1\times 10^{-5}$ -- $0.56$ & $\kappa_1=3.9\times 10^4$; $\kappa_2=8.4\times 10^3$ & $0.05$ -- $0.98$
\end{tabular}
\caption{Range of dimensionless parameters used in this study.\label{tab:dimless}}
\end{ruledtabular}
\end{table}

We should investigate the relations between global intensive variables, such as stresses, density, strain rate, in the limit of large samples, i.e., of $N\to+\infty$.  It is expected that for large enough samples the material state in shear flow will not depend on the specificities of boundary conditions, or on whether shear stress or strain rate is controlled. This requires the investigation of possible size effects and the study of the regression of fluctuations for global variables. Measured state variables should also be uniform in space -- and thus one needs to check for possible shear localization. If dimensionless variables such as stress ratios or density are well behaved in the triple limit of $N\to\infty$ (thermodynamic limit),
$I\to 0$ (quasistatic limit) and $\kappa\to+\infty$ (rigid limit), then the observed inner states and mechanical behavior of the packings
only depend on their geometric properties -- hence the name \emph{macroscopic geometric limit} we adopted for such a situation. One of the major
goals of the present study is the investigation of material properties in this limit.

Finally, as a practical application of the dimensional analysis of simulation parameters, let us note that the computational cost, expressed as a number of MD integration steps needed to reach a given shear strain $\gamma$, is proportional to $\gamma \sqrt{\kappa}/I$.

\section{Global variables and macroscopic behavior}\label{sec:global}
Our global observations and measurements are reported in this section. Conditions for proper observations of the intrinsic behavior of the material subjected to procedure D (shear-rate-controlled numerical experiments) are checked for in Section~\ref{sec:observations}, in which
various qualitative aspects of the material state in shear flow are discussed. Attention is then focused on macroscopic friction (Sec.~\ref{sec:macrofric}) and dilatancy (Sec.~\ref{sec:dilatancy}) properties of the material, which are more thoroughly and quantitatively investigated. Finally, the results obtained with procedure D at low inertial numbers are compared to those of the static approach, procedure S, in Section~\ref{sec:macrostatic}. Section~\ref{sec:macrodisc} discusses the essential results and their connections with the literature on granular materials.
\begin{figure}
\centering
\includegraphics[width = 8.5cm]{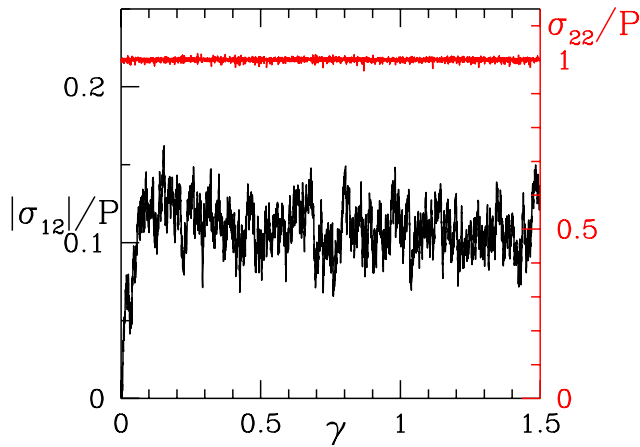}
\caption{(Color online) $|\sigma_{12}|$ (left axis, in black) and $\sigma_{22}$ (right axis, in red) as functions of strain $\gamma$. Note that the left and right scales are different. Time series obtained with $I = 3.2\times 10^{-5}$, $\kappa = \ka$, $\zeta = 0.98$ and $N = 4000$. \label{fig:transitoire}}
\end{figure}

\subsection{Material state in slow shear flow: qualitative aspects}\label{sec:observations}
With procedure D, we investigate steady states, and time series are collected for averaging. We are interested in intrinsic constitutive laws, as measured on averaging over the whole sample. It is therefore necessary to check for both invariance in time and homogeneity, in the statistical sense. We  should also assess the control of constant stress $\sigma_{22}$, and discuss the values of other stress components.
\subsubsection{Steady state flows and stress  measurements}
Fig.~\ref{fig:transitoire} displays the evolution of two components of the stress tensor, $\sigma_{22}$ and $\sigma_{12}$, with strain $\gamma$. It shows that $\sigma_{22}$ is well controlled since it was requested to stay equal to $\Sigma_{22}=0.1\, P$ in this numerical experiment. The evolution of stress $\sigma_{12}$, from the initial, isotropically confined state, witnesses the existence of an initial transient, which has virtually ended at $\gamma=0.1$ in that case. The steady state part of the time series starts for values of strain $\gamma$ that depend on the inertial parameter, of order $10^{-1}$ for the smallest $I$ values, ($\sim 10^{-5}$), increasing typically to about $0.5$ for $I=10^{-2}$ and to several units for $I\sim 10^{-1}$. Unlike in dense systems with intergranular friction~\cite{TH00,SUFL04,TER06}, for which deviator stresses, starting from isotropically compressed initial states,  go through a peak before approaching a plateau value at large strain, the shear stress in frictionless bead packs appears to grow monotonically, as a function of strain, toward its steady state value. Another notable feature of the shear stress as a function of time is the importance of fluctuations, which often exceed 30\% of the mean value on the example of Fig.~\ref{fig:transitoire}, in a sample of 4000 beads. A proper evaluation of average shear stresses thus requires careful statistical approaches and error estimates.

As a practical criterion to detect the end of the initial transient regime, we request that a small set of basic measured quantities do not exhibit any visible trend. Specifically, shear stress $\sigma_{12}$, volume fraction $\Phi$ and coordination number $z$ should all fluctuate about their mean value in a stationary manner, as well as the kinetic energy per particle, $\delta e_c$, associated with velocity fluctuations. The latter is defined as
\be
\delta e_c = \frac{1}{2N} \sum_{i=1}^N m \Big[(v_1 - \dot{\gamma}x_2)^2 + v_2^2 + v_3^2  \Big]
\label{eq:kinfluc}
\ee
$\delta e_c$ measures the instantaneous discrepancy between the actual flow generated by the Lees-Edwards
boundary condition in the granular material and the affine velocity field in a homogeneous continuum in shear flow.

Unlike $L_2$, lengths $L_1$ and $L_3$ are constrained to remain constant in procedure D, so that $\sigma_{11}$ and $\sigma_{33}$
may vary during the simulation.
For $I < 0.01$, we observed that time averages of $\sigma_{11}$ and $\sigma_{33}$ differed from the initial hydrostatic pressure $P$ by less
than $3\%$. This difference becomes even smaller for smaller inertial numbers:
for $I = 10^{-3}$, relative differences $(\ave{\sigma_{11}}/P)-1$ and $(\ave{\sigma_{33}}/P)-1$ respectively reduce to
$1.0\%$ and $2.2\%$. Those values decrease down to  $0.9\%$ and $1.7\%$ for
$I = 10^{-4}$, and to $0.6\%$ and $1.6\%$ for $I = 10^{-5}$. Although apparently not equal to zero,
even in the quasistatic limit, those stress components are very small, and, consequently, will not be studied in the sequel.
Sec.~\ref{sec:macrofric}, instead, focuses on accurate determinations of shear stress $\sigma_{12}$.

\begin{figure}[!htb]
\centering
\includegraphics[width=8cm]{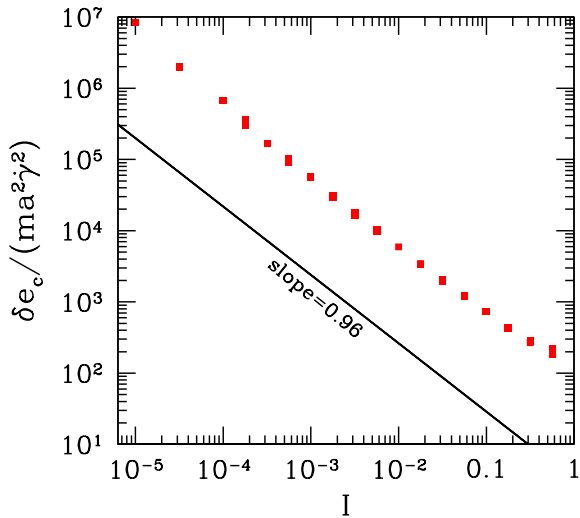}
\caption{Kinetic energy associated with velocity fluctuations, as defined in~\eqref{eq:kinfluc}, normalized by $ma^2\dot\gamma^2$, versus $I$, in simulations with
4000 beads, for $\kappa\in\{\kb,\ka \}$ and $\zeta = 0.98$.
\label{fig:kin4000}
}
\end{figure}

For a given number of particles, the \emph{relative} fluctuations of the instantaneous value of $\sigma_{12}$, $\Phi$ and $z$ (\emph{i.e.}, the ratio
of their quadratic average to the mean value) seem to be
independent of $I$. The average values of $\delta e_c$, on the other hand, as compared to the kinetic energy of the macroscopic field,
which is proportional to $\dot\gamma ^2$, increases as  $I$ decreases.
Fig.~\ref{fig:kin4000} is a plot of $\ave{\delta e_c}/(ma^2 \dot\gamma^2)$ versus $I$, showing that this ratio approximately diverges as $1/I$ in the limit
of $I\to 0$. This agrees with measurements made in 2D simulations of shear flows: the same behavior is reported in Ref.~\cite{Dacruz05}, and an interpretation was suggested,
to which we shall return in Section~\ref{sec:macrodisc}.
These observations suggest that in the quasistatic limit one has increasingly inhomogeneous instantaneous velocity gradient fields, which we now investigate.
\subsubsection{Instantaneous velocity profiles}
\begin{figure}
\centering
\includegraphics[width = 8.5cm]{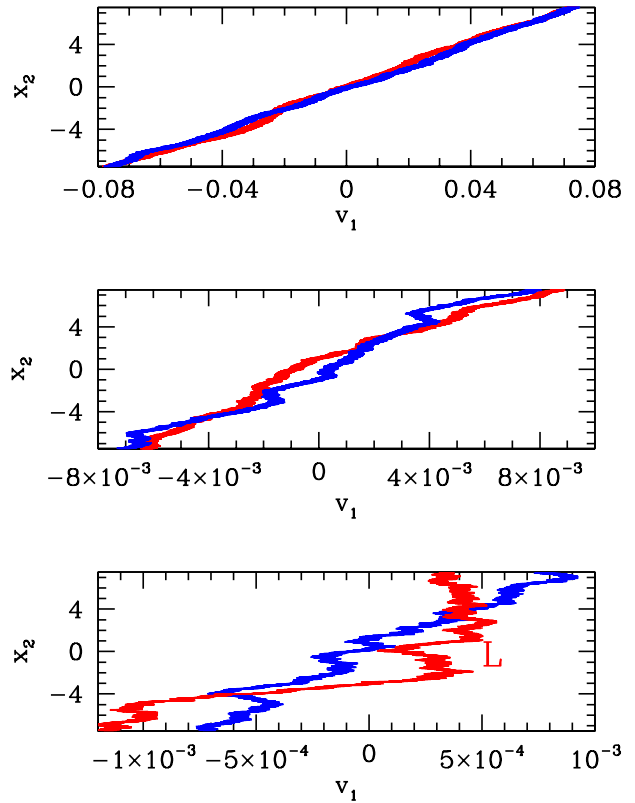}
\caption{(Color online)
Two velocity profiles at randomly chosen times, for $I = 3.2\times 10^{-2}$ (top), $I = 3.2\times 10^{-3}$ (middle), $I = 3.2\times 10^{-4}$ (bottom).
$\kappa = \ka$, $\zeta = 0.98$ and $N = 4000$. \label{fig:localisation}}
\end{figure}

Instantaneous velocity profiles $v_1(x_2)$
recorded at different random times for different values of $I$ are plotted in Fig.~\ref{fig:localisation}.
Profiles $v_1(x_2)$ are obtained on averaging particle velocities over slices cut alongside $x_2$ in the simulation cell (particles overlapping slice boundaries contribute to several different averages).
Inertial number $I$ has an important effect on the granular flow.
As shown in Fig.~\ref{fig:localisation}, instantaneous velocity profiles for $I = 3.2\times 10^{-2}$ are linear,
whereas shear bands may appear for $I = 3.2\times 10^{-4}$, as in the profile marked  ``L'' (for ``localized'') on the bottom plot of Fig.~\ref{fig:localisation}.
The transition between these two regimes seems to be gradual, with profiles in the middle part of Fig.~\ref{fig:localisation}, corresponding to $I = 3.2\times 10^{-3}$,
exhibiting somewhat intermediate features.

Localization occurs here in the bulk of the material since the system is not enclosed between walls.
Localization is thus an intrisic property of the studied material, which spontaneously appears for small values of $I$~\cite{AhSp02}.

\begin{figure}
\centering
\includegraphics[width = 8.5cm]{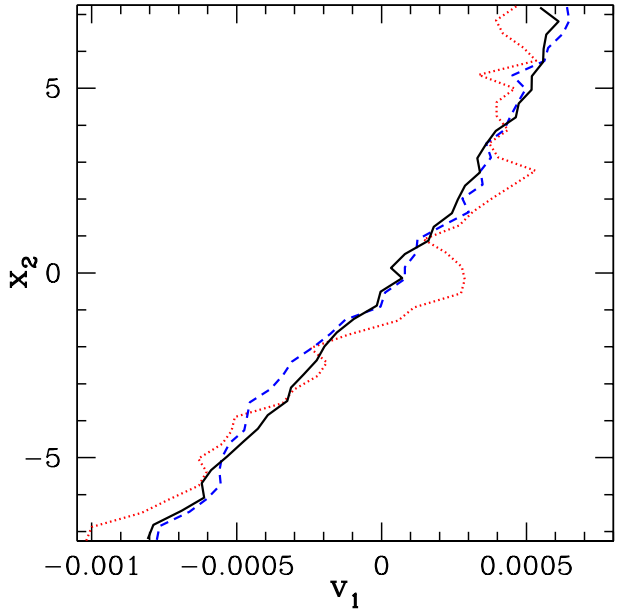}
\caption{(Color online) Velocity profile after shear strain intervals $\gamma$ equal to 0.004
(red dotted curve), 0.02 (blue dashed curve) and 0.1 (black solid curve), following the instant corresponding to
the localized profile marked L in Fig.~\ref{fig:localisation}, bottom graph.
\label{fig:profil}}
\end{figure}

At first glance, it seems that the erratic behavior of the velocity profiles in the quasistatic limit may
seriously jeopardize the interest of the results obtained on averaging over the whole simulation size and would demand specific analysis,
distinguishing between material states within and outside shear bands. However, localization patterns are not persistent, and linear velocity profiles
are  recovered by time averaging, even in the $I \rightarrow 0$ limit, which means that on average,
the flow is homogeneous. Figure~\ref{fig:profil} shows the gradual fading out of strain rate localization, after a strain interval of order $0.1$.
Shear bands thus randomly appear, move and disappear. Such a behavior is witnessed by larger relative fluctuations of $\delta e_c$ as $I$ decreases.  We did not carry out a detailed
study of the lifetime and dynamics of nonpersistent shear bands, as the statistical homogeneity of the system in steady state shear justifies an analysis
of global behavior based on averages over space and time.

\subsection{Macroscopic friction coefficient}\label{sec:macrofric}
For D simulations, the macroscopic friction coefficient, which we denote as $\mu^*$, is set equal to the time average -- in the steady state --
of the ratio of the shear stress $\sigma_{12} $ to the normal stress $\sigma_{22}$ (we use a convention where compressive stress components are positive)
\be
\label{eq:def_frott}
\mu^* =\langle\frac{|\sigma_{12}|}{\sigma_{22}}\rangle_t
\ee

The simulations produce raw data in the form of time series.
The steady part of the time series is isolated as explained in Sec.~\ref{sec:observations}
and $\mu^*$ can then be easily computed. To estimate the statistical uncertainty on the measurement of averages over finite time series, we use the ``blocking'' (or ``renormalization group'') technique
presented in~\cite{FP89}. This yields error bars on measurements of averages in finite systems
which should not be confused with the quadratic average of fluctuations of the observable quantity. In practice, due to intrinsic long-lasting correlations in the system, we observed that
quite long runs were necessary. In some cases with $I\sim 10^{-5}$, up to $10^9$ simulation time steps (corresponding to a deformation $\gamma \simeq 4$) were necessary for a correct evaluation of the
uncertainty on $\mu^*$.

In the present case, we could check that the time series of all observable quantities recorded in distinct samples differing only by their initial state were statistically identical in steady state with high accuracy, as expected from critical state theory~\cite{DMWood,RR04,RK04,KR06}.

$\mu^*$, as estimated from time series in type D simulations,
may depend on the three dimensionless numbers introduced in Sec.~\ref{sec:dimanalysis}, as well as on
the number $N$ of particles. This dependence is investigated in the following paragraphs.
\subsubsection{Effect of $I$}
\begin{figure}
\centering
\includegraphics[width = 8.5cm]{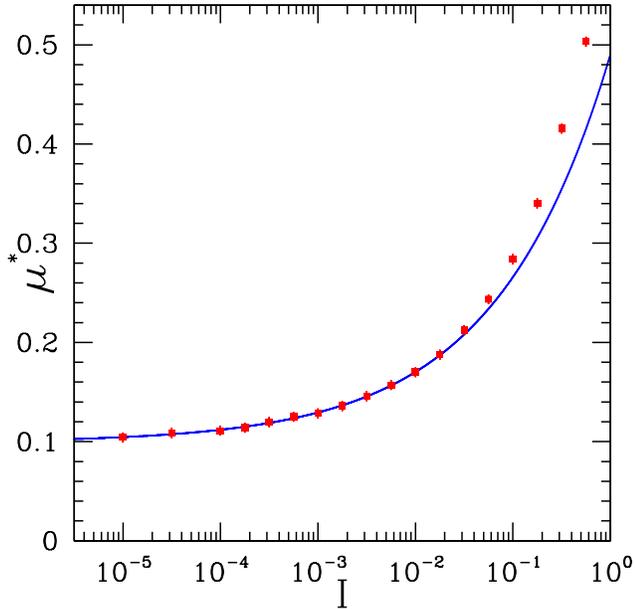}
\caption{Macroscopic friction $\mu^*$ \emph{vs.} inertial number $I$ for stiffness parameter
$\kappa\in\{\kb,\ka\}$, damping parameter $\zeta=0.98$ and number of beads $N = 4000$.
The solid line is Eq.~\eqref{eq:muIfit} with the parameters of Table~\ref{tab:mufitpar} (no visible difference on using best parameters with
$\ka$ or $\kb$).\label{fig:frott}}
\end{figure}

Among the three dimensionless parameters governing the behavior of the system, the inertial number $I$
has the strongest effect on $\mu^*$. Fig.~\ref{fig:frott} plots $\mu^*$ as a function of $I$. It shows that $\mu^*$ is an increasing function of $I$. This dependence of the macroscopic friction coefficient
on the inertial number is similar to the ones reported in the literature, as obtained by both simulations and experiments~\cite{Gdr04,Dacruz05,Hatano07}, although
most published results pertain to granular systems with friction in the contacts.
Here $\mu^*$ approaches a finite nonzero value $\mu_0^*$ in the quasistatic limit of $I\to 0$, despite the absence of friction at intergranular contacts.
$\mu_0^*$ coincides with the internal friction coefficient of the material in its critical state.

Note the accuracy of the displayed curve: statistical uncertainties measured with the blocking method are comprised between
$10^{-4}$ and $10^{-3}$ and are thus invisible on the graph.

\subsubsection{Effect of $\kappa$}
Near the rigid limit, the
macroscopic behavior should reflect the absence of stress scale in the contact law:
stress ratios and derived quantities such as the macroscopic friction coefficient are expected to be independent of the average stress.
The friction coefficient hardly changes between the two values of $\kappa$ used in our simulations, indicating that
the rigid limit $\kappa\to\infty$ is accurately approached.
Those simulations were carried out with $\zeta = 0.98$ for $1.8\times 10^{-4} \leq I \leq 5.6\times 10^{-1}$, and the
relative variation on $\mu^*$ is less than $2\%$ throughout this range of inertial parameter
on varying the stiffness parameter from $\kappa = \ka$ to $\kappa = \kb$.
Thus we decided to gather the values obtained for the macroscopic friction coefficient
with $\kappa = \kb$ and $\kappa = \ka$ in
Fig.~\ref{fig:frott}, because the uncertainty on the macroscopic, geometric limit of $\mu^*$ to be estimated will eventually exceed this small difference.

\subsubsection{Effect of $\zeta$\label{sec:zetaonmu}}
The viscous damping term is indispensable in the model, as the only source of dissipation, but its magnitude should be irrelevant in the quasistatic limit. Consequently, the influence of $\zeta$ on our results had to be assessed and we performed simulations for
different values of $I$ with $\zeta = 0.98$ (this value corresponds to a restitution coefficient
$e_N = 3.3\times 10^{-3}$), $\zeta = 0.55$ ($e_N = 0.17$), $\zeta = 0.25$ ($e_N = 0.49$) and $\zeta = 0.05$ ($e_N = 0.87$).
Our results show that for $I < 10^{-3}$, the maximal relative variation of $\zeta$ on the macroscopic
friction $\mu^*$ is less than $1\%$. Far from the quasistatic regime, the influence of $\zeta$ is no more negligible:
the relative variation of $\mu^*$ is greater than $10\%$ on changing $\zeta$ when $I > 10^{-1}$.

\subsubsection{Effect of $N$\label{sec:Nonmu}}
The influence of the sample size on the average of the apparent friction coefficient $|\sigma_{12}| / \sigma_{22}$,
was investigated on
comparing results for three different numbers of particles: $N = 500$, $N = 1372$ and $N = 4000$.
Results are listed in Table~\ref{tab:Nonmu}. We also recorded the standard deviations, denoted as $\Delta \mu$, and
the average of the top percentile of the instantaneous values, denoted as $\mu^{*,100}$.
Let us recall that we are dealing here with the fluctuations of the time series, which  differ from the statistical uncertainties on the average values.
\begin{table}
\begin{ruledtabular}
\begin{tabular}{cccccccc}
$I$ & $N$ & $\mu^*$ & $\Delta\mu / \mu^*$ & $\mu^{*,100}$ & $\Phi$ & $\Delta\Phi / \Phi$ & $\Phi^{,100}$ \\ \hline
\rule{0mm}{4.5mm}\multirow{3}{*}{$3.2 \times 10^{-5}$} &  500 & 0.1169 & 0.3100 & 0.2188 & 0.6367 & 0.0022 & 0.6403 \\
                                                       & 1372 & 0.1101 & 0.1907 & 0.1609 & 0.6380 & 0.0015 & 0.6408 \\
                                                       & 4000 & 0.1090 & 0.1245 & 0.1431 & 0.6387 & 0.0008 & 0.6404 \\
\rule{0mm}{4.5mm}\multirow{3}{*}{$3.2 \times 10^{-4}$} &  500\footnotemark[1] & 0.1432 & 1.263 & 0.8378 & 0.6738 & 0.0178 & 0.7148 \\
                                                       & 1372 & 0.1209 & 0.1689 & 0.1779 & 0.6365 & 0.0016 & 0.6390 \\
                                                       & 4000 & 0.1197 & 0.1002 & 0.1519 & 0.6368 & 0.0010 & 0.6388 \\
\rule{0mm}{4.5mm}\multirow{3}{*}{$3.2 \times 10^{-3}$} &  500 & 0.1473 & 0.2091 & 0.2275 & 0.6316 & 0.0027 & 0.6360 \\
                                                       & 1372 & 0.1457 & 0.1293 & 0.1966 & 0.6322 & 0.0016 & 0.6346 \\
                                                       & 4000 & 0.1458 & 0.0764 & 0.1752 & 0.6323 & 0.0009 & 0.6338 \\
\rule{0mm}{4.5mm}\multirow{3}{*}{$3.2 \times 10^{-2}$} &  500 & 0.2112 & 0.2045 & 0.3317 & 0.6193 & 0.0025 & 0.6234 \\
                                                       & 1372 & 0.2123 & 0.1197 & 0.2802 & 0.6197 & 0.0015 & 0.6223 \\
                                                       & 4000 & 0.2125 & 0.0694 & 0.2517 & 0.6200 & 0.0009 & 0.6215\\
\end{tabular}
\caption{Influence of sample size $N$ on macroscopic friction $\mu^*$ and volume fraction $\Phi$
for different values of inertial number $I$, with $\kappa=\ka$ and $\zeta=0.98$.
Superscripts ``$,100$'' denote the average of the top percentile values in the steady state part of the time series.
\label{tab:Nonmu}}
\end{ruledtabular}
\footnotetext[1]{This numerical experiment displays shear-induced ordering, a feature observed only for systems of $N < 1000$ beads (see Appendix~\ref{sec:appcryst} for details).}
\end{table}

The effect of the sample size on the macroscopic friction is unnoticeable for $N = 1372$ and $N = 4000$
since the difference between the friction coefficients pertaining to these two sizes is less than the statistical
uncertainty marring the accuracy on $\mu^*$. However, the impact of $N$ on $\mu^*$ cannot be neglected in the quasistatic limit for a system of $N = 500$ beads.
These results show that some minimum number of beads, of order about 1000, is required to approach the thermodynamic limit with an acceptable accuracy.

\begin{figure}
\centering
\includegraphics[width = 8.5cm]{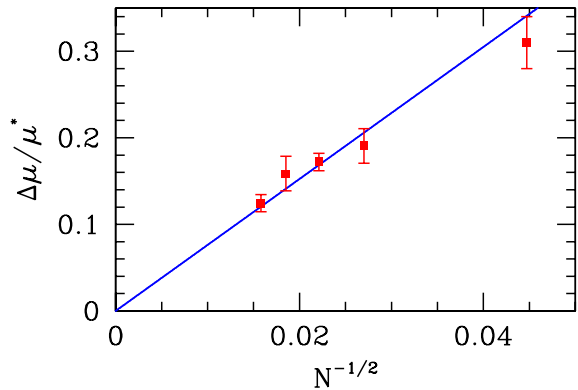}
\caption{$\Delta\mu / \mu^*$ as a function of $N$ for $I = 3.2\times 10^{-5}$,
$\kappa = \ka$ and $\zeta = 0.98$. The solid line equation is relation~\eqref{eq:flucrelmu}.
\label{fig:flucrelmu}}
\end{figure}

The data of Table~\ref{tab:Nonmu} also witness the regression of fluctuations of stress ratio $\mu^*$ in the steady state in the large system limit.
The results are compatible with the classical form for the decrease of fluctuations of collective variables, \emph{viz.} $\Delta \mu/\mu^* \propto N^{-1/2}$.
Specifically, for $I = 3.2\times 10^{-5}$,
$\kappa = \ka$ and $\zeta = 0.98$, a fit of the data to the following form:
\be
\Delta \mu/\mu^* = (7.6 \pm 0.7) N^{-1/2}
\label{eq:flucrelmu}
\ee
has good statistical admissibility.
This result is shown in Fig.~\ref{fig:flucrelmu} in graphical form (two additional sizes $N=2048$ and $N=2916$ were also simulated).
Therefore, we expect the steady state stress-strain curves such as the ones of Fig.~\ref{fig:transitoire},
however noisy for the sample sizes simulated, to become smooth
in the large system limit.

\subsubsection{Approach to the macroscopic geometric limit}\label{sec:macrofricgeom}
According to the previous parametric study, the geometric limit can be confidently studied as the limit of
$I \rightarrow 0$ on samples of $4000$ beads with $\kappa \geq \kb$ and $\zeta = 0.98$.

$\mu^*$ should be a function of the sole inertial number in very good approximation for sufficiently small values of $I$.
In the absence of any scale, we tried to fit the data by a power law function (see Fig.~\ref{fig:frott}) of the form
\be
\mu^*=\mu_0^* + cI^{\alpha}
\label{eq:muIfit}
\ee
As stated above, this fit is not expected to be accurate for large $I$ values and
we therefore restricted ourselves to fit the data points with $I \leq 0.01$. Parameters $\mu^*_0$ (the geometric macroscopic friction coefficient),  $\alpha$ and $c$ were separately estimated for $\kappa=\ka$ and $\kappa=\kb$ (keeping $\zeta=0.98$ and $N=4000$) and results are shown in Table~\ref{tab:mufitpar}.
\begin{table}
\begin{ruledtabular}
\begin{tabular}{cccc}
$\kappa$ & $\mu_0^*$ & $\alpha$ & $c$  \\ \hline
$\ka$ & $0.101\pm 0.00$4 & $0.38\pm 0.04$ & $0.40\pm 0.07$ \\
$\kb$ & $0.100\pm 0.003$ & $0.39\pm 0.02$& $0.42\pm 0.03$\\
\end{tabular}
\caption{Best fit parameters for Eq.~\eqref{eq:muIfit} and the data obtained with $N=4000$, $\zeta=0.98$ for $\kappa=\ka$ and $\kappa=\kb$.
\label{tab:mufitpar}
}
\end{ruledtabular}
\end{table}

The value of the geometric macroscopic friction angle $\varphi_0^*$ corresponding to $\mu^*_0$ is
(for $\kappa=\ka$)
\be
\varphi_0^* = 5.76^{\circ} \pm 0.22^{\circ}
\label{eq:valphi0}
\ee
Quite similar values are also reported with two-dimensional packings of polydisperse disks by Taboada~\emph{et al.}~\cite{TER06},
whose estimate of the macroscopic friction angle lies between $4^{\circ}$ and $7^{\circ}$ for frictionless grains, and by Da Cruz~\emph{et al.}~\cite{Dacruz05},
who obtained $\mu^*_0\simeq 0.1$ in shear flow simulations for small $I$ parameters.
Hatano~\cite{Hatano07} recently performed 3D numerical simulations on polydisperse assemblies of about 10000 spherical beads, for different intergranular
friction coefficients $\mu$. The reported value of the macroscopic friction coefficient in the quasistatic limit is $0.06$ for $\mu=0$, apparently lower than our result.
It should be recalled however that Hatano's work was motivated by applications to
granular materials under high confining stresses within geological fault zones, and
that consequently simulations were carried out with lower stiffness levels ($\kappa = 1840$, $136$, $84$ and $42$) than in the present study. Moreover,
the lower range of $I$ parameters, below $3\times 10^{-4}$, was only investigated with the lower $\kappa$ values.
Hatano used the same form as Eq.~\eqref{eq:muIfit} to fit his data,
but his estimate $\alpha = 0.28\pm 0.05$ differs from ours (see Table~\ref{tab:mufitpar}). Although some effect of the polydispersity is possible, we also
attribute this discrepancy to some non-negligible influence of $\kappa$ in Hatano's simulations~\cite{Hatano07}.
Only the simulations with $\kappa = 1840$ in~\cite{Hatano07} can be expected to approach the rigid limit accurately.
For this stiffness level, Hatano's data points are available for $I \geq 3\times 10^{-4}$
and are in very good agreement with ours (e.g., $\mu^*\simeq 0.17$ for $I=0.01$).

\subsection{Dilatancy and steady-state density\label{sec:dilatancy}}
Dilatancy under shear is a basic property of
granular materials in quasistatic deformation~\cite{RE85,DMWood,PWR62,TER06}, when dense samples are subjected to increasing deviator stresses.
The steady-state density is mainly sensitive to $I$ if $\kappa$ is large enough~\cite{Dacruz05}. The small $I$ behavior of frictionless bead assemblies is investigated here with greater accuracy than in previous studies.

We could check that, just like the macroscopic friction coefficient, the steady state time average of the volume fraction, $\Phi \equiv \ave{\Phi(t)}_t$,
is an intrinsic property of the material, independent of its initial preparation.
Next, we investigate its dependence on the three dimensionless parameters $I$, $\kappa$ and $\zeta$ and on the number of particles $N$.

\subsubsection{Effect of $I$, $\kappa$ and $\zeta$}
\begin{figure}
\includegraphics[width = 8.5cm]{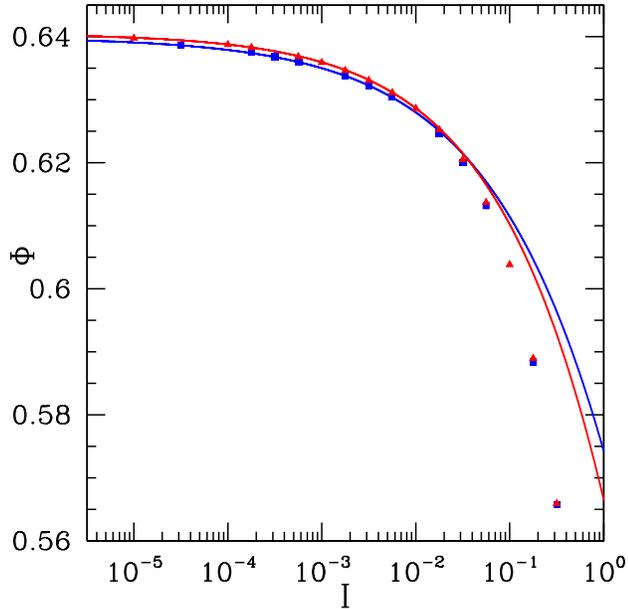}
\caption{(Color online) Volume fraction $\Phi$ as a function of inertial number $I$ (for $\zeta = 0.98$, $N = 4000$),
for both stiffness levels $\kappa=\ka$ (blue squares) and $\kappa=\kb$ (red triangles)
The solid lines are given by Eq.~\eqref{eq:phiIfit} with the parameters of Table~\ref{tab:phifitpar}.
\label{fig:dilat}}
\end{figure}

Once again, numerical experiments demonstrate that among the three dimensionless numbers governing the behavior of the system,
the inertial number $I$ has the most important effect on $\Phi \equiv \ave{\Phi(t)}_t$.
Fig.~\ref{fig:dilat} shows the influence of $I$ on $\Phi$. We observe that $\Phi$ decreases for increasing $I$, as previously reported~\cite{Dacruz05,Hatano07}.
It starts from a value $\Phi_0\simeq 0.64$ in the quasistatic limit and the system expands as $I$ increases.
Statistical uncertainties on $\Phi$ measured thanks to the blocking method are comprised between $10^{-5}$ and $10^{-4}$ and are thus invisible on the figure.
$\Phi_0$ is very close to $\Phi_{\text{RCP}}$~\cite{OSLN03,iviso1}, which coincides (up to
small corrections due to the finite contact stiffness)
with the initial volume fraction $\Phi^{\textrm{iso}}$, right after
the samples are assembled with procedure O. For
$\kappa = \kb$ and $N = 4000$ we have $\Phi^{\textrm{iso}} = 0.6382\pm 0.0011$, and
$\Phi^{\textrm{iso}} = 0.6369\pm 0.0009$ for $\kappa = \ka$ (averages and standard deviations are evaluated on five samples).
The system studied thus appears to be devoid of dilatancy in the quasistatic limit. Whether $\Phi_0$ should be regarded as equal to
$\Phi^{\textrm{iso}}\simeq\Phi_{\text{RCP}}$ at the macroscopic level
will be discussed after the possible influence of $N$ on the average densities is investigated.

Stiffness parameter $\kappa$ typically induces a relative increase of the volume fraction of roughly $0.1\%$
when it changes from $\kappa = \kb$ to $\kappa = \ka$, whatever the value of $I$ --
a small effect, yet distinguishable from statistical uncertainties. Such a density increase is of course expected, as
larger contact deflections due to larger stresses or a softer material decrease the sample volume.
Simulations with $\zeta = 0.98$ ($e_N = 3.3\times 10^{-3}$), $\zeta = 0.55$ ($e_N = 0.17$),
$\zeta = 0.25$ ($e_N = 0.49$) and $\zeta = 0.05$ ($e_N = 0.87$) for a wide range of inertial numbers have also been run.
The influence of $\zeta$ on $\Phi$ is important for large $I$: for $I > 0.1$, the relative variation of $\Phi$ with $\zeta$ can reach $30\%$.
However, this effect, as expected, gradually vanishes as the quasistatic limit is approached, and for $I < 0.01$
the relative variation of $\Phi$ with $\zeta$ is less than $0.1\%$.
\subsubsection{Effect of $N$\label{Nonphi}}
\begin{figure}
\centering
\includegraphics[width = 8.5cm]{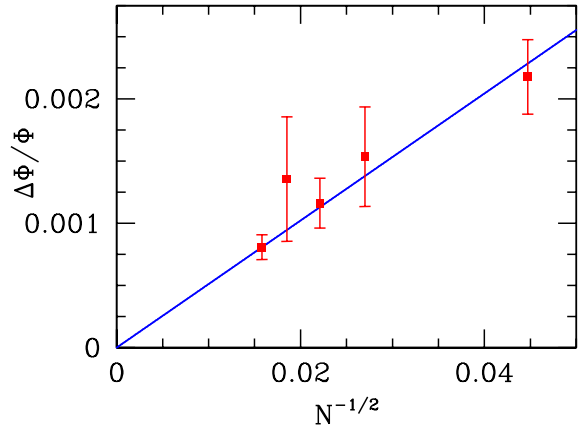}
\caption{$\Delta\Phi / \Phi$ as a function of $N$ for the same time series as in Fig.~\ref{fig:flucrelmu},
fitted with Eq.~\eqref{eq:flucrelphi} (solid line).
\label{fig:flucrelphi}}
\end{figure}

According to Table~\ref{tab:Nonmu}, $\Phi$ very slightly varies with the number $N$ of particles, like in static, isotropic systems~\cite{OSLN03,iviso1}.
The following fit, based on the measurements for the smallest available value of $I$, i.e., $I=I_1=3.2\times 10^{-5}$ for
$\kappa=\ka$, may be used:
\be
\Phi (\kappa= \ka, I=I_1,N)  = \Phi_1 - k_1 N^{-1/2}
\label{eq:fitphiN}
\ee
with the following parameters:
\be
 \left\{ \begin{array}{ll}
\Phi_1 &= 0.6398\pm 0.0002\\
k_1 &=0.070\pm 0.008\\
         \end{array}
 \right.
\label{eq:phiparfitN}
\ee
As with the macroscopic friction coefficient $\mu^*$, we could check for the regression of fluctuations of the volume fraction for increasing $N$.
For the same set simulations with $I = 3.2\times 10^{-5}$,
$\kappa = \ka$ and $\zeta = 0.98$ as in Sec.~\ref{sec:Nonmu}, we observe that the decrease of density fluctuations with increasing $N$ can be fitted by
the following relation:
\be
\frac{\Delta\Phi}{\Phi}= \frac{A}{\sqrt{N}},\ \ \mbox{with $A=0.051\pm 0.011$}
\label{eq:flucrelphi}
\ee
as shown graphically in Fig.~\ref{fig:flucrelphi},
whence a well-defined large system limit for $\Phi$.
\subsubsection{Approach to the macroscopic geometric limit}
Volume fraction $\Phi$ can therefore be modeled as a function of $I$ near the quasistatic limit,
say for $I < 0.01$, with small corrections to account for the influence of $N$ and $\kappa$.
It can be regarded as independent of $\zeta$ (at least for $I < 0.01$).
The fit form used is
\be
\Phi^{-1} = \Phi_0^{-1} + e I^{\nu}
\label{eq:phiIfit}
\ee
For $N=4000$ and $\zeta = 0.98$, the best fit values of the parameters of Eq.~\eqref{eq:phiIfit} are given in Table~\ref{tab:phifitpar}.
\begin{table}
\begin{ruledtabular}
\begin{tabular}{cccc}
$\kappa$ & $\Phi_0$ & $\nu$ & $e$  \\ \hline
$\ka$ & $0.6398\pm 2.\,10^{-4}$ & $0.39\pm 0.01$ & $0.1786\pm 8.\,10^{-4}$ \\
$\kb$ & $0.6405\pm 2.\,10^{-4} $& $0.42\pm 0.02$& $0.2038\pm 3\,10^{-4}$\\
\end{tabular}
\caption{Best fit parameters for Eq.~\eqref{eq:phiIfit} and the data obtained with $N=4000$, $\zeta=0.98$ for $\kappa=\ka$ and $\kappa=\kb$.
\label{tab:phifitpar}
}
\end{ruledtabular}
\end{table}

To evaluate the macroscopic value $\Phi_0^*$ in the double limit of $I\to 0$ and $N\to +\infty$,
it is reasonable to assume that the small corrections to $\Phi_0^*$ that result from
the finite value of $N$ and from the nonvanishing value of $I$ are additive. The use of Eqs.~\eqref{eq:fitphiN}-\eqref{eq:phiparfitN} to evaluate
the finite $N$ correction to the value $\Phi_0$ of the quasistatic density, as obtained on fitting Eq.~\eqref{eq:phiIfit} to the results with $N=4000$,
yields, for $\kappa=\ka$:
\be
\Phi_0^*= 0.6410 \pm 0.0005
\label{eq:phi0macro}
\ee
The increases of $\Phi$, from its value in the rigid limit, due to the finite stiffness is of order $\kappa^{-1}$ (see~\cite[Eq. 31]{iviso1})
and is smaller than the statistical uncertainty in~\eqref{eq:phi0macro}.
The value of $\Phi_0^*$ given in~\eqref{eq:phi0macro} is thus our best estimate, from D-type simulations, of the solid fraction of sheared sphere packings in the
macroscopic geometric limit.
\subsection{Static behavior\label{sec:macrostatic}}
We now compare the results of Sections~\ref{sec:macrofric} and \ref{sec:dilatancy}
for steady shear-rate-controlled simulations (procedure D) with those obtained through static shear numerical experiments (procedure S).

\subsubsection{Friction coefficient\label{sec:macrofricstat}}
The static macroscopic friction coefficient is defined in procedure S as
\be
\mustat = \frac{|\tau_{\textrm{max}}|}{P}
\ee
where $\tau_{\textrm{max}}$ denotes the maximum shear stress which the system has been able to sustain in mechanical equilibrium, and $P$ the confining pressure.
Static microscopic friction coefficients for different sample sizes are displayed in Table~\ref{tab:frott_stat}.
\begin{table}
\begin{ruledtabular}
\begin{tabular}{cccc}
N & $S_N$ & $\langle \mustat \rangle$ & $\Delta \mustat$ \\
\hline
256 & 4 & 0.246 & 0.022 \\
500 & 4 & 0.210 & 0.007 \\
1372 & 6 & 0.169 & 0.004 \\
2048 & 6 & 0.154 & 0.004 \\
2916 & 6 & 0.145 & 0.007 \\
4000 & 10 & 0.136 & 0.007 \\
8788 & 6 & 0.122 & 0.005 \\
\end{tabular}
\end{ruledtabular}
\caption{\label{tab:frott_stat} Average $\langle \mustat \rangle$
and standard deviation $\Delta \mustat$ of the static friction coefficient obtained in S-type simulations, over $S_N$ samples of
$N$ grains for different $N$. Data corresponding to both values of $\kappa$ (with $S_N/2$ samples each) are aggregated.}
\end{table}
Values of $\mustat$ are larger than the dynamical value $\mu^*_0 = 0.100 \pm 0.004$ obtained in D simulations in the quasistatic limit.
As shown by Tab.~\ref{tab:frott_stat}, $\mustat$ is size dependent, unlike $\mu^*_0$ (for $N \gtrsim 1000$).
Analogous observations were reported in~\cite{XO06} for two-dimensional systems of frictionless disks:
in the limit of vanishing shear rates, the shear stress reaches its large system limit with only several hundreds of beads,
whereas the minimum shear stress required to maintain a long lasting steady shear flow exceeds the previous one and is more sensitive to $N$.

\begin{figure}
\centering
\includegraphics[width = 8.5cm]{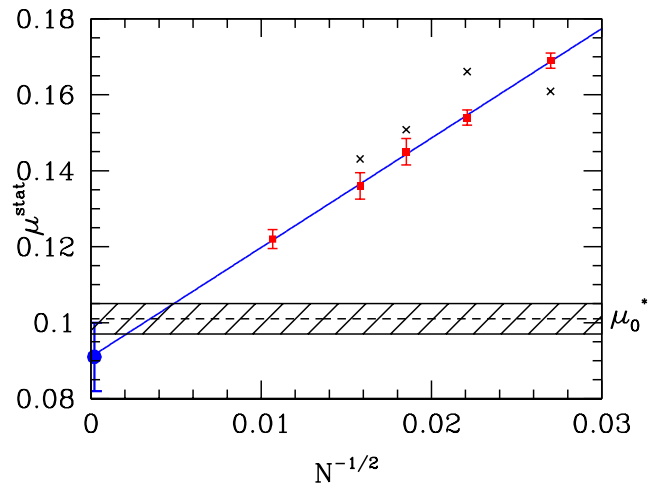}
\caption{(Color online) Size dependence of $\ave{\mustat}$. $N$ denotes the number of particles in the system.
The solid line is the fit of Eqs.~\eqref{eq:mustatNfit}-\eqref{eq:mustatNfitpar}.
Crosses are the top percentile values extracted from the time series of $|\sigma_{12}| / \sigma_{22}$ obtained in procedure D,
as listed in Table~\ref{tab:Nonmu}. The hashed region represents the estimate, from D simulations, of $\mu^*_0$ with its error bar
(Table~\ref{tab:mufitpar}). The (blue) dot with an error bar on the left axis is the static estimate, $\mustatinf$.\label{fig:musize}}
\end{figure}

Fig.~\ref{fig:musize} shows the influence of system size $N$ on $\mustat$ (discarding the smallest $N$ values).
The data are correctly fitted by the following relation
\be
\mustat = \mustatinf + dN^{-1/2},
\label{eq:mustatNfit}
\ee
with
\be
 \left\{ \begin{array}{lll}
             \mustatinf& = & 0.091 \pm 0.009 \\
             d  & = & 2.87\pm 0.32
         \end{array}
 \right.
\label{eq:mustatNfitpar}
\ee
The related angle of friction is $\varphi_{\infty}^{\textrm{stat}} = 5.20^{\circ} \pm 0.52^{\circ}$.
This is consistent with the equality, in the thermodynamic limit ($N \rightarrow \infty$), of the dynamical and static macroscopic friction coefficients (see Eq.~\ref{eq:valphi0}).
The influence of $\kappa$ is very small and negligible in comparison to the effect of the system size, and we therefore averaged over systems
with both stiffness levels $\ka$ and $\kb$.

With the smallest system size simulated, $N=256$, we observed that some of the samples, once submitted to shear stresses, acquired a strongly
ordered, crystalline structure, to be discussed in Appendix~\ref{sec:appcryst}.
\subsubsection{Density\label{sec:dilatstat}}
\begin{figure}
\includegraphics[width = 8.5cm]{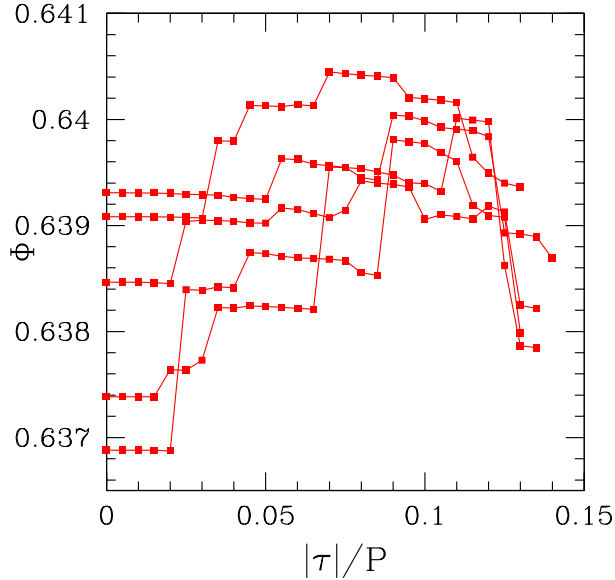}
\caption{Variation of the volume fraction $\Phi$ with the static shear $|\tau| / P$ imposed to five different samples of
4000 beads with $\kappa = \kb$. Each curve stops at a given value of $|\tau| / P = |\tau_{\textrm{max}}| / P$:
this is the greatest value for which the packing has managed to reach mechanical equilibrium.\label{fig:comp-frott}}
\end{figure}

Static shear simulations with procedure S support the observation made in Sec.~\ref{sec:dilatancy}
that the frictionless model material studied is devoid of dilatancy in the quasistatic limit.
As shown in Fig.~\ref{fig:comp-frott}, which represents $\Phi$ as a function of the macroscopic stress ratio $\tau / P$ imposed to the material in different samples
with $N=4000$,
the volume fraction hardly evolves with the stress deviator as it is increased towards its maximum value. However, whatever the initial state of the system, it experiences a slight compaction at the beginning of the shear and a small decompaction near the failure limit, but we have no convincing explanation for this phenomena. The evolution of $\Phi$
is somewhat erratic (as in previous studies on 2D rigid, frictionless disk assemblies~\cite{CR2000,RC02})
and the density change between the isotropic initial state and the one supporting the maximum shear stress
is equal to zero, within statistical uncertainties.
Similarly to the values of $\Phi$ measured in D simulations, solid fraction $\phistat$ in static packings under maximum shear stress is slightly
dependent on sample size, with a negative finite-size correction to the macroscopic value. On fitting a variation proportional to $N^{-1/2}$ one gets,
for $\kappa=\ka$,
\be
\phistat = \phistatinf - k / \sqrt N,
\label{eq:phistatNfit}
\ee
with
\be
 \left\{ \begin{array}{lll}
             \phistatinf& = &0.6403 \pm 0.0004\\
             k  & = & 0.125 \pm 0.026.
         \end{array}
 \right.
\label{eq:phistatNfitpar}
\ee
$\Phi$ values for  $\kappa=\kb$ are slightly larger, by about $10^{-3}$.

Comparing this estimate of $\phistatinf$ with the result for $\Phi_0^*$ given in~\eqref{eq:phi0macro}, we conclude that
static and dynamic solid fractions in quasistatic shear are identical, within statistical uncertainties.
Disregarding the very small correction due to the finite value of $\ka$ (equal to about $1.1\times 10^{-4}$
on applying the formula given in \cite[Eq. 31]{iviso1}), this means that,
just like for $\mu^*$, the values of
solid fraction $\Phi$ in the macroscopic, geometric limit coincide in strain rate controlled and in shear stress controlled approaches.

As to the value $\phiso$ of the solid fraction in the initial isotropic state,
a similar evaluation of size effects yields (using the samples of Table~\ref{tab:frott_stat}
with $\kappa=\ka$ and $N\ge 500$):
\be
\phiso = \phisoinf -\frac{k_0}{\sqrt{N}},
\label{eq:phiNiso}
\ee
with
\be
 \left\{ \begin{array}{lll}
             \phisoinf& = &0.6397 \pm 0.0008\\
             k_0  & = & 0.15 \pm 0.03.
         \end{array}
 \right.
\label{eq:phiNisofitpar}
\ee
As announced, this is the random close packing value~\cite{OSLN03,iviso1}.
Results~\eqref{eq:phiNisofitpar}, \eqref{eq:phistatNfitpar} and \eqref{eq:phi0macro} are compatible,
and we thus conclude that the system is devoid of dilatancy under shear in the macroscopic geometric limit.
\subsection{Discussion\label{sec:macrodisc}}
We briefly review and comment here the essential results on the macroscopic behavior of the material under simple shear, and compare them to
other available results on similar systems.
\subsubsection{Internal friction and the macroscopic geometric limit}
Whether assemblies of frictionless grains have a well-defined,
finite internal friction coefficient has sometimes appeared as a debatable issue, although some previously
cited works~\cite{Dacruz05,Hatano07,OSCS98} relying on numerical simulations of slow shear flows in steady state agree with our positive conclusion. A proper evaluation of $\mu^*_0$ in the macroscopic geometric limit requires more care than corresponding measurements in granular assemblies with friction. This is due to the importance of fluctuations, as apparent on Fig.~\ref{fig:transitoire}. In D-type simulations, it is also necessary to explore a range of very small inertial numbers to accurately evaluate the quasistatic friction coefficient,
as apparent in Fig.~\ref{fig:frott}. As an example, for $I=5.6\times 10^{-4}$, quite a small value,
the macroscopic friction coefficient already exceeds its quasistatic limit by 25\%.

Our estimate for $\mu^*_0$ is confirmed by static simulations, once the results are suitably extrapolated to the limit of large systems. One may understand this size effect on S results as follows. The friction coefficient evaluated in D simulations is an average over time series with large fluctuations. However, the system remains close to mechanical equilibrium at any time. Assuming it is possible to find an equilibrated configuration very close to all dynamically explored states, however large the instantaneous value of the shear stress, the S procedure would be able to find statically supported shear stress values as large as the maximum of $\sigma_{12}$ in D time series. Although clearly oversimplifying the evolution of the system in configuration space, this explanation appears to be correct at least on correlating $N$-dependent maximum static shear stress levels to fluctuations in slow shear flows: the $N$ dependence of $\mustat$, as plotted in Fig.~\ref{fig:musize}, is paralleled by that of the typical largest values of $\mu^*$ (top percentile) in D simulations.

Static and dynamic values of shear stress thresholds for flow are also observed to coincide in the fixed density simulation results of Xu and O'Hern~\cite{XO06}, obtained on 2D packings of frictionless disks, with a similar excess of the static value that vanishes as $N\to\infty$.

When non negligible inertial effects are present, we observe that the increase of internal friction with inertial number $I$ is the dominant feature of the material behavior (the effect of stiffness level $\kappa$ is smaller by orders of magnitude), in qualitative agreement with many other results on frictional and frictionless grains~\cite{Dacruz05,Hatano07}.
\subsubsection{Absence of dilatancy}
Our results also agree in static shear and in steady state constant shear rate flows for the average volume fraction $\Phi$, which stays
equal to its value in the initial, isotropically confined configuration in the macroscopic geometric limit. Our data show that, within statistical uncertainties (i.e., about $5\times 10^{-4}$) the critical value of $\Phi$ is equal to $\Phi_{\text{RCP}}$ in packings of frictionless spherical beads.

The material studied is thus devoid of dilatancy. Interestingly, this contradicts the simple pictures of the origins of dilatancy which have been proposed since the introduction of this property by Reynolds~\cite{RE85}, based on the distortion of simple assemblies of a small number of contacting spheres (like, e.g., a regular tetrahedron)~\cite{GODI98}.

The absence of dilatancy in the quasistatic limit is also at odds with the classical ideas on the relation between dilatancy and internal friction, according to which macroscopic friction stems from two microscopic origins, intergranular friction and dilatancy, with an additive combination of relevant angles~\cite{PWR62,GODI98}. Ref.~\cite{TER06} adds another component  $\varphi_0$ to macroscopic friction, due to intergranular collisions as a source of dissipation, and therefore accounts for the internal friction of frictionless grains. Thus $\varphi_0$ is the internal friction angle that we measure in the geometric limit. Ref~\cite{TER06}, although only incidentally dealing with frictionless materials, nevertheless appears to predict a positive dilatancy in that case, which our results do not confirm. Similarly, a recent study published by Kruyt and Rothenburg~\cite{KR06}, which also deals with 2D disk assemblies, predicts a non-vanishing dilatancy when intergranular friction coefficient $\mu_0$ approaches zero.  Ref.~\cite{KR06}, similarly to Ref.~\cite{TER06}, discusses stress-dilatancy relations, and finds a linear variation of the dilatancy ratio with the difference between peak and steady-state macroscopic friction. In contradiction with our data, it attributes a positive value to both quantities as the friction coefficient approaches $0$, while its estimate for $\mu^*_0$ is significantly larger than our (3D) one, or than the (2D) one of Ref.~\cite{TER06}. (Note that the maximum deviator to mean stress ratio, as defined in~\cite{KR06,RR04}, is $\sin\varphi \simeq \tan\varphi$). The frictionless case was not directly simulated in this work. Some rapid variations of macroscopic friction and dilatancy angles near the singular limit of $\mu_0 \to 0$ might be overlooked.

In our simulations, instantaneous fluctuating shear stress and volume fraction, however, appear to be correlated, suggesting some stress-dilatancy coupling at the level of short-lived, transient and rearranging structures, which disappears on taking time averages.

\subsubsection{A toy model \label{sec:toy}}
Since some of our results on the macroscopic behavior of frictionless bead packs might seem counter-intuitive,
we designed a simple model in which similar basic ingredients (geometric constraints defining isolated equilibrium positions,
inertia, viscous dissipation) produce an analogous behavior in a suitably defined
``macroscopic geometric limit''. In both cases, the microscopic motion is a succession of arrested
dynamical phases, alternating with approaches to transient equilibria.  The toy model simply provides suggestive analogies,
it should not be regarded as a real physical explanation for the macroscopic behavior of the granular system.

\begin{figure}[!htb]
\centering
\includegraphics[width=8cm]{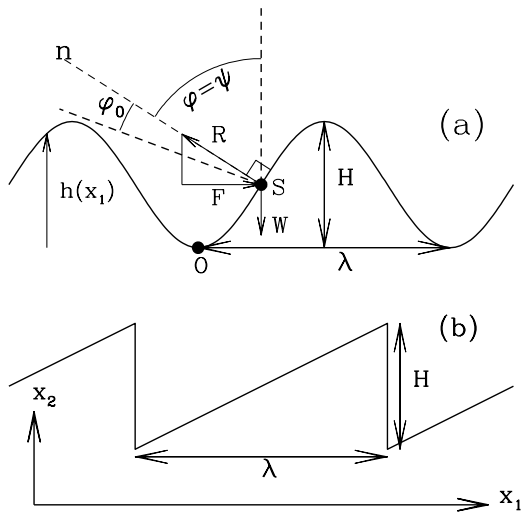}
\caption{The model of the slider on a rough surface. (a) Case of a sinusoidal profile. Forces at point S are drawn as vectors. (b) Profile for which $\mu_D=\mu_S$.
\label{fig:slider}}
\end{figure}

We consider  a single object of mass $M$, subject to its weight $W$, pushed along a rough horizontal surface.
For simplicity the model is two-dimensional, with only one horizontal coordinate, $x_1$, and the surface profile $h(x_1)$, along vertical coordinate $x_2$,
is periodic with wavelength $\lambda$, as depicted in Fig.~\ref{fig:slider}. The mobile object
is driven either by a constant horizontal force $F$, or by a piston with constant horizontal velocity $V$. Both contacts are rigid and unilateral,
so that the mobile object might move faster than the piston if accelerated downhill by gravity, or
occasionally take off from the surface.
Force $F$ is the analog of shear stress $\sigma_{12}$ in the granular material,
and $W$ that of $\sigma_{22}$, while horizontal and vertical displacements respectively correspond to shear strain and volume increase.
A viscous force opposes the tangential motion along the surface, so that for $F=0$
the slider stabilizes at some local minimum of profile $h(x_1)$, like point O on the figure. Such minima are analogous to equilibrium states under isotropic pressure.

Let us first discuss the static experiment, in which the mobile object, initially in equilibrium in $O$ under $F=0$,
is subjected to a growing horizontal force $F$. It equilibrates where the tangent direction
to the substrate is orthogonal to the applied force, $\frac{dh}{dx}=F/W$. It has first to move upwards, hence some \emph{dilatancy}. The maximum value
of $F/W$ is the static effective friction coefficient $\mu_S=\tan\varphi$ of the object on the surface, equal to the maximum slope of profile $h(x_1)$. It is reached at point S on the
figure. Effective \emph{static} friction angle $\varphi$ is the maximum angle between the reaction of the substrate, force ${\bf R}$ on Fig.~\ref{fig:slider}, and the vertical
direction.
As the quasistatic motion from O to S follows the surface, dilatancy $\tan\psi$,
defined as the ratio of vertical to horizontal coordinates of the velocity (corresponding to ratio $\dot\epsilon_{22}/\dot \gamma$ in the sheared
granular material), is also identical to the maximum profile slope. Dilatancy and friction
angles coincide: $\psi=\varphi$. If a nonzero friction coefficient $\mu_0 = \tan\varphi_0$ is introduced
in the contact between the mobile object and the substrate, then reaction ${\bf R}$ at point $S$ (see Fig.~\ref{fig:slider}) may form an angle $\varphi_0$ with normal
direction $(Sn)$, so that the effective static friction angle is $\varphi = \varphi_0+\psi$ -- a classical form of the
\emph{stress-dilatancy relation}~\cite{DMWood,TER06}.

In the velocity-controlled case, the \emph{dynamic} friction coefficient is conveniently evaluated from the dissipation of energy. In the limit of small
velocity, the mobile object pulls ahead of the velocity-controlled driving piston at each maximum of $h(x_1)$.
Its subsequent downhill sliding is accelerated by gravity, but it is
prevented by viscous dissipation to pass the next maximum, and ends up at the bottom of the valley,
where it is later picked up by the slow piston, to be pushed up the next ascending slope.
In this scenario the dissipated energy per wavelength $\lambda$ is the potential energy loss $HW$  in a fall over height $H$.
Hence an effective friction coefficient $\mu_D= H/\lambda$. This result is, remarkably, independent of the viscous damping coefficient,
just like the macroscopic friction of  the granular material in slow shear flow. As the properties of the system only depend then on substrate geometry, the limit of slow imposed
velocity is the \emph{geometric limit}.

The \emph{macroscopic limit} can be defined as $\lambda/L \to 0$, where $L$ is the length scale on which the effective properties of the slider
are studied. Consequently, its vertical motion, on the scale $H\sim\lambda$ of microscopic asperities of the surface, becomes irrelevant, and
one observes \emph{effective macroscopic friction without dilatancy}. Models for dilatancy~\cite{GODI98} apparently
focus on microscopic phases of the motion in which the slider rises up the slope, but ignore the equally important ones in which it falls down.

$\mu_D$ is the \emph{average} slope of the ascending part of the profile,
multiplied by the fraction of length for which $h(x_1)$ is increasing. It is in general smaller than $\mu_S$, which is the \emph{maximum} slope.
Thus, for a sinusoidal profile, $h(x_1) = H/2 \sin (2\pi x_1/\lambda)$, as represented on Fig.~\ref{fig:slider}(a),
one has $\mu_S = \pi H/\lambda$, while $\mu_D = H/\lambda$.  In order for both friction coefficients
to coincide, function $h(x_1)$ should be as shown on Fig.~\ref{fig:slider}(b), with ascending parts of constant slope, followed by vertical drops.

The particular profile shape of Fig.~\ref{fig:slider}(b) can be argued to
be appropriate for the analogy with the bulk material. As long as the contact with the substrate is maintained, the configuration might be an
equilibrium position for some (possibly negative) value of $F$. In the analogy with the granular material, the contact network might balance the external load for
some value of the applied stress components. The free fall, on the other hand, is the analog of a network rearrangement, during which
applied loads cannot be supported  because of the missing contacts. In the granular material (as explicitly shown in~\cite{CR2000}) intervals of stress components for which
a given contact network is stable shrink to zero in the large system limit. This corresponds for the toy model
to a constant slope of the rising parts of profile $h(x_1)$ (defining a unique possible value of $F$ in equilibrium).

Finally, the velocity-controlled sliding of the object on the profile of Fig.~\ref{fig:slider}(b)
also provides an interpretation of inertial number $I$~\cite{Gdr04},
and of the behavior of the kinetic energy~\cite{Dacruz05}. The motion involves two characteristic times: the duration of the rising phase,
in which the object is in contact with the piston and moves with horizontal velocity $V$, $\tau_1 = \lambda/V$; and the duration of the free fall,
$\tau_2 \propto \sqrt{(MH/W)}$. Their ratio, $\tau_2/\tau_1 \propto (V/\lambda)\sqrt{(MH/W)}$, is the analog of number $I$, as readily checked on replacing
distance $H$ by some length of order $a$ (a typical interstice between neighboring grains to be closed for a new contact to appear),
$V/\lambda$ by $\dot\gamma$, and force $W$ by $Pa^2$, which is the order of magnitude of unbalanced forces on the grains in the dynamical phases of motion.
The free fall phases of the motion explain why the
kinetic energy is, on average, much larger than the scale $MV^2$ associated with the macroscopic motion. More precisely, the time average $\delta e_c$
of the kinetic energy associated with velocity fluctuations is of order $HW (\tau_2/\tau_1)$ (for $\tau_2\ll\tau_1$), whence (discarding
constant factor $(H/\lambda)^2$) the behavior shown in Fig.~\ref{fig:kin4000}, $\delta e_c\propto (MV^2)/I$.

\section{Microstructure and force networks\label{sec:micro}}
Our specific emphasis on the \emph{geometric limit} of the macroscopic mechanical behavior of frictionless bead packings calls for an analysis of the geometry
of sheared configurations, the first motivation of which is to explain the microscopic origins of macroscopic friction.
Ultimately, a model should be sought which, unlike
the analogous one of Sec.~\ref{sec:toy}, would explicitly and quantitatively describe the mechanisms,
involving instabilities and network rearrangements at the microscopic level,
by which the material deforms and flows. Such goals will be only partly achieved here, since,
leaving the detailed study of velocity correlations and strain mechanisms for
future work, we focus on simple characterizations that are local in space and time.
We also check here that the various microstructural variables studied, if measured in D-type simulations, approach their values
observed in S-type ones in the limit of $I\to 0$ (at least in the large system limit).

Packing geometry is classically described with a few state variables, among which the simplest ones are scalar:
the volume fraction, the coordination number, as studied here in Section~\ref{sec:conn} below.

The much-studied distribution of contact force values~\cite{RJMR96,BMMJN01} is also determined in the present case (Section~\ref{sec:forces}),
and we check for effects of inertia and anisotropy.

Under stress, or influenced by the history of their assembling process, the microstructure of grain packings develops anisotropic features, which are most
often characterized with the \emph{fabric tensors}, expressing statistics on orientations of normal directions at contacts,
as studied in Sec.~\ref{sec:fabric}. The critical state is microscopically characterized by stationary values of $\Phi$, $z$, and fabric tensors, which are reached after a sufficiently large interval of monotonically growing strain
in the quasistatic regime~\cite{RTR04,RR04,RK04}.

\subsection{Coordination number\label{sec:conn}}

\begin{figure}
\includegraphics[width = 8.5cm]{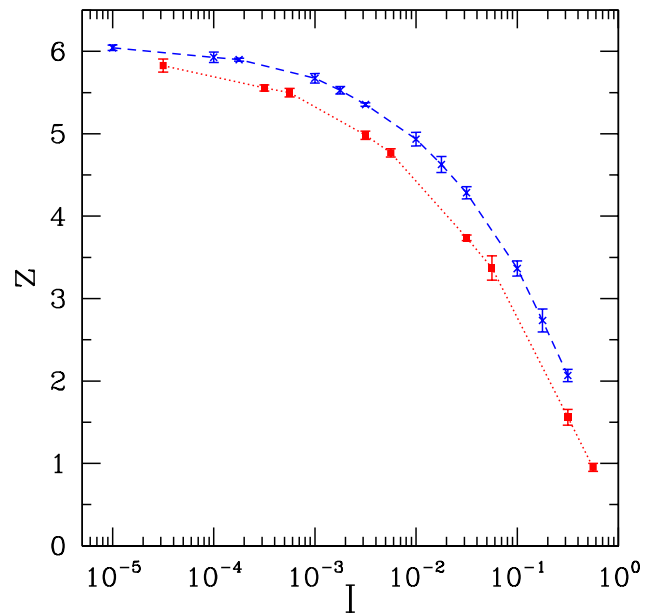}
\caption{\label{fig:z}(Color online) Coordination number $z$ as a function of inertial number $I$, for $\kappa = \ka$
(red square dots joined by dotted line, bottom data points) and $\kappa = \kb$ (blue crosses, top, dashed line).}
\end{figure}

The coordination number $z$ strongly depends on $I$ in steady state shear flow, and it is also affected by $\kappa$. It decreases with increasing $I$, or with increasing $\kappa$. As to the influence of $\zeta$ on $z$, it is notable for the largest $I$ values explored, but it decreases as the quasistatic limit is approached. Larger viscous damping coefficients increase the duration of contacts in shear flow, and thus produce
slightly better coordinated networks on average. However, the intensity of viscous forces becomes irrelevant in the quasistatic limit. According to our results, the $\zeta$ dependence of $z$ can safely be ignored for $I < 10^{-4}$. The $I$ dependence of $z$ is shown in Fig.~\ref{fig:z}.

The coordination number of the equilibrated (S-type) anisotropic configurations is also very close to 6. This is a consequence of the isostaticity property of the force-carrying structure (also called \emph{backbone}~\cite{iviso1}) of equilibrated sphere packings in the rigid limit -- a remarkable property discussed in several recent publications~\cite{OSLN03,DTS05,iviso1}, which is specific to packings of rigid, frictionless and cohesionless spherical grains~\cite{JNR2000,Donev-ellipse}.

Fig.~\ref{fig:z} shows that for quite low values of $I$, many contacts are lost ($z$ is down to about 5 for $I$ in the $10^{-3}$ range). The proportion $p_0$ of rattlers increases with increasing $I$: $p_0$ is less than $1.5\%$ in S simulations and for D simulations with $I \leq 10^{-4}$ and $\kappa = \ka$, but is equal to $30\%$ for $I = 3.2\times 10^{-1}$ and $\kappa = \ka$. Our results are compatible with the theoretical value $z = 6(1-p_0)$ in the limit $I \to 0$ and $\kappa \to +\infty$. Furthermore, it has been often observed that the grains only have a small number of contacts bearing large forces -- this is the very reason why  the ``force chains'' exist~\cite{OR97b,RWJM98,SRSvHvS05}. Consequently, as contacts carrying smaller forces are necessarily shorter lived, and tend to rarefy as $I$ increases, the populations of grains with the largest local coordination are quickly depleted.

\subsection{Distribution of forces\label{sec:forces}}
\begin{figure}
\includegraphics[width = 8.5cm]{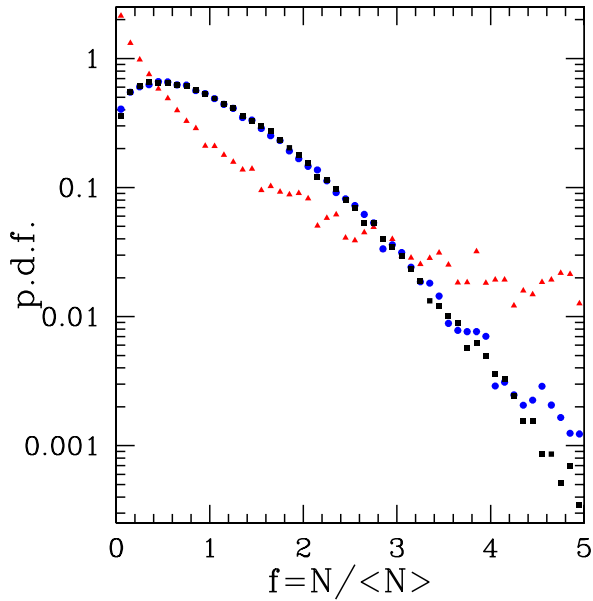}
\caption{(Color online) Probability distribution function of $f = N/\langle N \rangle$ for $I = 3.2\,10^{-1}$ (red triangles),
$I = 3.2\,10^{-3}$ (blue round dots) and $I = 3.2\,10^{-5}$ (black square dots). \label{fig:fntot}}
\end{figure}

Fig.~\ref{fig:fntot} is a plot of the probability distribution function of the intergranular force normalized by the average force,
for different values of $I$.
The force distribution strongly depends on $I$:
for $I > 3.2\times 10^{-2}$ the probability distribution function $p(f)$ ($f$ denoting the ratio of the normal force to
the average value $\ave{F_N}$) is monotonically decreasing.
For smaller values of $I$, $p(f)$ has a maximum, around $f=0.5$, and an approximately exponential decay for large values, as often
observed in equilibrated granular packings~\cite{CLMNW96,RJMR96,OR97b,BMMJN01,SGL02,DTS05,iviso1}.
The distributions obtained for the low values of $I$ in D-type simulations gradually approach the one obtained in S-type,
equilibrated packings under maximum shear stress.
The Kolmogorov-Smirnov test~\cite{numrec} can be used to detect the influence of parameters on the force distribution -- the answer depending of course
on the level of statistics of the available data. Based on 10 independent configurations of 4000 grains, it leads to the conclusions that no
significant difference in force distribution could be detected between S-type results under maximum shear stress and D-type ones, and no influence of
$\kappa$ either, provided the inertial parameter is small enough: $I < 5\times 10^{-3}$, while some influence of $\zeta$ is only visible for $I>0.1$. Our results are also compatible with a unique distribution, valid for maximum shear stress equilibrium configurations as well as for isotropic ones.
\subsection{Fabric\label{sec:fabric}}
Macroscopic friction is known to stem (at least partially) from the build-up of fabric anisotropy
in materials made of frictional beads or disks~\cite{RR04,RTR04}. This connection is explored here with frictionless beads.

Anisotropy of the tridimensional contact network can be characterized by the probability density function
$E(\theta,\varphi)$ of finding a contact with direction $(\theta,\varphi)$. $\theta$ is the colatitude angle and $\varphi$ is the longitude angle of the spherical coordinates.
$E$ can be expanded in a series of
spherical harmonics. The coefficients of the expansion
are in one-to-one correspondence with the values of the fabric tensors, which are defined as the moments
of the distribution of normal unit vectors $\vec{n}$ on the unit sphere.
Since a contact is left invariant by the parity symmetry $\vec{r}\rightarrow -\vec{r}$, $E$ satisfies
$E(\theta,\Phi) = E(\pi - \theta,\varphi + \pi)$. This means that the coefficients of odd order
in the expansion in spherical harmonics are all equal to zero, and
corresponds to the vanishing of all odd order fabric tensors. Coefficients can be computed from even order tensor products, \emph{viz.}
\be
\langle \bigotimes_{i=1}^{2k} \vec{n} \rangle \equiv \frac{1}{N_c}\; \sum_{c\in\mathcal{C}}\;
\bigotimes_{i=1}^{2k} \vec{n}_c
\ee
$\mathcal{C}$ denoting the set of $N_c$ contacts, labeled with index $c\in\mathcal{C}$, where the normal unit vector is $\vec{n}_c$.

Keeping only the lowest order of anisotropy, the expansion of $E$ is restricted to the spherical harmonics of order two.
Coefficients of the development are directly related to the value of the fabric tensor of order two, denoted by $\ww{F}$:
\bea
E(\theta,\varphi) & = & 1/(4\pi) + F_{12} d_{xy}(\theta,\varphi) \nonumber \\&+ &(F_{11} - F_{22}) d_{x^2-y^2}(\theta,\varphi) \nonumber \\
 & + & (F_{33} - 1/3) d_{z^2}(\theta,\varphi) + F_{13} d_{xz}(\theta,\varphi) \nonumber \\
 & + & F_{23} d_{yz}(\theta,\varphi) + \textrm{higher order terms} \label{eq:sph_harm_decomp}
\eea
The constant $1/(4\pi)$ corresponds to an isotropic distribution and the next five terms of the
development characterize the anisotropy of the material at the lowest order.
Functions $d$ are combinations of spherical harmonics of order two, with following expressions:
\begin{eqnarray}
  d_{xy}(\theta,\varphi) &=& \frac{15}{8\pi}\sin^2 \theta \sin(2\varphi) \\
  d_{x^2 - y^2}(\theta,\varphi) &=& \frac{15}{16\pi}\sin^2 \theta \cos(2\varphi) \\
  d_{z^2}(\theta,\varphi) &=& \frac{15}{16\pi}(3\cos^2 \theta - 1) \\
  d_{xz}(\theta,\varphi) &=& \frac{15}{4\pi} \sin\theta\cos\theta\cos\varphi \\
  d_{yz}(\theta,\varphi) &=& \frac{15}{4\pi} \sin\theta\cos\theta\sin\varphi
\end{eqnarray}

\begin{figure}
\includegraphics[width = 8.5cm]{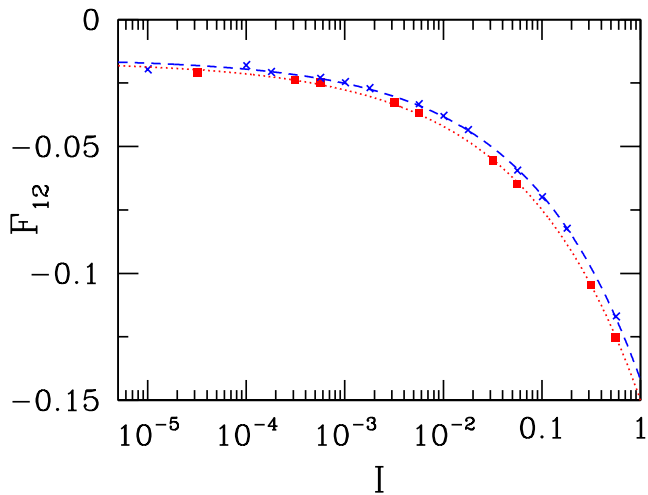}
\caption{ (Color online) $F_{12}$ as a function of inertial number $I$, with $\zeta = 0.98$ and $N = 4000$, for $\kappa = \ka$ (red square dots connected by a dotted line) and for $\kappa = \kb$ (blue crosses connected by a dashed line). Both lines are power law fits of $F_{12}$.
\label{fig:fabric}}
\end{figure}

Fabric tensor $\ww{F}$ is computed as a time average in the steady shear simulations.
The numerical results show that $F_{13}$ and $F_{23}$ are always less than their respective statistical uncertainties, and
can be considered as equal to zero, as requested by the symmetry in simple shear.
We observe that $F_{12}$ is always greater (by at least one order of magnitude)
than the two other non negligible anisotropic coefficients, $F_{11} - F_{22}$ and $F_{33}-1/3$.
These two latter terms are below $2\times 10^{-3}$ for $I < 10^{-3}$. Such low values
are comparable with sample to sample fluctuations in  equilibrated configurations.
Thus, in the quasistatic limit, the anisotropy can be characterized by the sole $F_{12}$ coefficient, the limit of which,  as $I\to 0$,
is evaluated as $F_{12}^0=-0.0165\pm 7.\, 10^{-4}$ for $\kappa=\ka$ and  $F_{12}^0=-0.0156\pm 7.\, 10^{-4}$ for $\kappa=\kb$, with a fitting procedure. Like $\mu^*$
and $\Phi$, $F_{12}$ strongly varies with $I$ (Fig.~\ref{fig:fabric}), and
its dependence on $I$ can be represented by a power law, with exponent $\simeq 0.36$ for $\kappa = \ka$ and $\simeq 0.37$ for $\kappa = \kb$.

$F_{12}$ values measured in S-type equilibrated samples under maximum shear stress are influenced by system size $N$, very similarly to the static friction
coefficient: larger values of $\vert F_{12}\vert$ are observed (typically $\simeq 0.02$ for $N=4000$),
but the excess over
$F_{12}^0$, the estimate from D-type simulations in
the quasistatic limit, regresses as $N$ increases, and the extrapolated macroscopic limit is compatible with the estimated values of $F_{12}^0$.
This will be further examined in the more general context of the relationship between stress and anisotropies, for arbitrary stress tensors, in a forthcoming publication~\cite{PEPJNR08b}.

On changing $\zeta$ from $0.98$ down to $0.05$,  $\vert F_{12}\vert$ increases (correlatively with the decrease of $z$), by about 30\% for $I\sim 10^{-2}$. This relative change is reduced to about 1\% for $I\sim 10^{-4}$ and the effect of $\zeta$ vanishes in the limit of $I\to 0$.

Variations of $F_{12}$ with parameters $I$, $\kappa$ and $\zeta$ are qualitatively understood on noting that $F_{12}$ is negatively
correlated with the coordination number. If there are more contacting neighbors, on average, around a sphere, they are prevented by steric constraints from
achieving highly anisotropic orientation distributions. This argument, which with simple assumptions was made quantitative in 2D in Ref.~\cite{RTR04}, thus explains
that the increase of $z$ observed as $\kappa$ is lowered tends to reduce $\vert F_{12}\vert$, as observed on Fig.~\ref{fig:fabric}. Similarly, the larger anisotropies observed away from the quasistatic limit
are made possible by the smaller number of contacts. The increase of $|F_{12}|$ with $I$ is also
due to the correlation of force intensities with contact directions: on evaluating
separately the fabric of the subnetworks corresponding to forces larger (or smaller)
than the average contact force, one typically obtains, for $I\sim 10^{-5}$,  values of
$\vert F_{12}\vert$ twice as large (respectively: four times as small) as with the complete contact network.
Contacts with small forces open if $I$ is increased, and the remaining more strongly loaded ones are consequently more anisotropically oriented.

\section{Discussion\label{sec:dis}}
This work was devoted to the study of frictionless identical spherical balls subjected to simple shear. The influence of the three dimensionless quantities controlling the problem -- inertial number $I$, stiffness number $\kappa$ and level of viscous damping $\zeta$ -- was carefully assessed and we observed that $I$ has the most dramatic impact on the system behavior. Fluctuations of the measured quantities were shown to vanish for large systems. Consequently, the particular nature of the boundary conditions employed has no importance: for sufficiently large systems, fixed-volume simulations would lead to the very results we obtained with our stress driven numerical experiments. Particular attention was paid to the macroscopic geometric limit, that is the triple limit $N\rightarrow +\infty$, $I \rightarrow 0$ and $\kappa\rightarrow +\infty$. In this r\'egime, the system behavior is governed by a succession of instabilities due to dynamical rearrangements of the contact network. A thorough investigation of such events remains an interesting, yet challenging, perspective.

The existence of a nonzero macroscopic friction angle was evidenced by two different kinds of simulations -- shear-rate controlled dynamic calculations (D-type simulations) and quasistatic stress-controlled calculations (S-type simulations). Whereas the dynamic friction angle $\varphi_D$ is independent of the system size for $N > 1000$, the static friction angle $\varphi_S$ is very sensitive to the number of grains and is systematically greater than $\varphi_D$ for all studied sizes ($N\leq 8788$) and $\varphi_S - \varphi_D$ increases for decreasing $N$. This might be the reason why localization seems to occur more easily as the system size decreases. In finite-size systems, the shear stress is a multivaluated function of the strain rate in the quasistatic limit and the range of multivaluation increases with decreasing $N$. Thus shear bands are more likely to appear in small systems~\cite{Dacruz05,VBBB03,VBB04}. However, in the macroscopic geometric limit, we found that both friction angles $\varphi_S$ and $\varphi_D$ are equal within statistical uncertainties. In frictionless granular assemblies, all dissipation is due to viscous terms in contact forces, which therefore can be regarded~\cite{TER06} as the physical origin of macroscopic friction. However, the value of the damping coefficient $\zeta$ is irrelevant in the quasistatic limit since the amount of dissipated energy is geometrically determined. In the macroscopic geometric limit, we have seen that the shear has no effect on the microscopic scalar quantities of the material (coordination number, distribution of forces), but it induces some structural anisotropy and a correlation between force intensities and contact orientations. We thus attribute the macroscopic friction angle to the shear-induced anisotropy of the material, as in the frictional case~\cite{ARPS07}. Ref.~\cite{PEPJNR08b} will show quantitatively that this is indeed the case.

The result that $\varphi_S = \varphi_D$ contrasts with observations on Lennard-Jones glasses at temperature $T > 0$~\cite{VBBB03,VBB04} and on granular avalanches~\cite{Pouliquen99,DaerrDouady99}. Glass simulations show that the dynamic angle is less than the static one. This difference is linked to a stress overshoot visible on strain-stress curves. Similarly, in dense granular materials with friction, the shear stress goes through a maximum before the steady state (``critical state") is reached, a feature which is absent in frictionless granular assemblies (both states coincide in this case). Similar differences ($\varphi_S > \varphi_D$) are reported for granular flows down inclined plane. Thus, in Refs.~\cite{Pouliquen99,DaerrDouady99},  $\theta_{\textrm{stop}}(h)$ is less than $\theta_{\textrm{start}}(h)$, where $\theta$ is the inclination of the plane and $h$ the thickness of the flowing layer in the stationary state. The small thickness of the layer (typically less than ten grain diameters) and the intergranular friction are certainly responsible for this hysteresis.

The stress-dilatancy interplay is a well known feature of granular materials. However, our simulations show that homogeneously sheared frictionless bead assemblies do not display any dilatancy in the macroscopic geometric limit. In this limit, volume fraction $\Phi$ remains equal to $\Phi_{\textrm{RCP}}$ during the whole time the material is sheared and the backbone stays isostatic in the rigid and quasistatic limits. This surprising lack of dilatancy can be intuitively understood in the light of the simple model presented in Sec.~\ref{sec:toy}. We thus conclude that the steady state (critical) volume fraction $\Phi_{\textrm{c}}$ is equal to $\Phi_{\textrm{RCP}}$.

The behavior of frictionless granular assemblies under arbitrary load directions will be the subject of a future work~\cite{PEPJNR08b} in order to gain a better knowledge of the yield surface and of the mechanical properties of such granular systems under a small enough stress deviator (before failure).

One motivation of the present work is the study of highly concentrated non-Brownian suspensions (P\'eclet number Pe $=+\infty$), modeled as assemblies of nearly touching grains bonded by a viscous lubricant~\cite{MB95,MB97,MB04b}. Ideal lubrication effectively suppresses the tangent forces. Lubricated dynamics has already been employed as a means to obtain the force-carrying contact network of frictionless rigid particles, as the set of viscous bonds on which stresses concentrate~\cite{OR95}. Although crude, our current model should be able to reproduce the behavior of dense suspensions in the quasistatic limit. In this r\'egime, the system evolves \emph{via} a sequence of equilibrium states. At some point, the initial network is no longer able to sustain the imposed stress, it becomes unstable and a dynamic ``crisis" occurs. Consequently, the evolution of the system is not quasistatic in the strictest sense (each point of the configuration space cannot be reached through a continuous series of equilibrium configurations). However, details of the dynamics are expected to be irrelevant. Thus we expect that the same equilibrium states will be visited in the quasistatic limit by both frictionless granular systems and dense suspensions with frictional grains.
According to the simple toy model of Sec.~\ref{sec:toy}, a dense suspension might be sketched by a slider moving on a bumpy surface in a media of viscosity $\eta$. Close to the quasistatic limit, the most important parameter would be the dimensionless number $\eta\dot{\gamma}/P$. One may notice that it is very similar to the parameter $I_v$ introduced by Cassar \emph{et al.} that controls submarine avalanches in what they call the viscous regime~\cite{CNP05}. Steady shear simulations evidenced that the material is still able to flow with a volume fraction approximately equal to $\Phi^* = \Phi_{\textrm{RCP}} \simeq 0.64$. This result is consistent with theoretical results pertaining to suspensions, where the volume fraction $\Phi^*$ at which the viscosity of the suspension diverges is believed to tally with the random close packing volume fraction~\cite{SB01}. However, it is not in agreement with the experiments exposed in~\cite{OBR06}, where the value of $\Phi^*$ was found to be below $0.61$. This discrepancy very likely originates in small scale features of the experimental system that are not accounted for a model of perfectly lubricated spherical beads. The behavior of dense suspensions is known to be strongly impacted by short-range physics~\cite{MB04a}. In the near future, we plan to study lubricated pastes with frictional contacts in the spirit of the simplified Stokesian dynamics scheme proposed by Ball and Melrose~\cite{MB95,MB97,MB04b}.

\appendix
\section{Crystallization under shear\label{sec:appcryst}}
Small samples, in both D and S-type simulations, tend to form strongly ordered structures under shear.
This phenomenon, which do not occur for $N > 1000$, is briefly reported here. A more detailed study would be outside the scope of
the present paper, and would require some investigation of the role of cell shape and boundary conditions, which is necessarily important
in such small systems.

2 out of 3 S-type samples with $N=256$ and stiffness level $\kb$, and 2 out of 2 D-type samples with  $N=500$, $I = 3.2\times 10^{-4}$ and $\kappa=\ka$
present the following anomalies. First, solid fractions are considerably higher than $\Phi_{\text{RCP}}$ (and even more so considering the
size effect~\cite{OSLN03,iviso1} on $\Phi$), with values approximatively equal to $0.67$ (see fourth line of Tab.~\ref{tab:Nonmu}). Apparent friction coefficients are also particularly large. A lower bound for the static macroscopic friction coefficient of S-type ordered samples is $0.4$, whereas dynamic macroscopic friction coefficient $\mu^*$ of D-type ordered samples for $I = 3.2\times 10^{-4}$, $\kappa = \ka$, and $N = 500$ may exceed by $20\%$ the corresponding friction coefficient in bigger samples that do not experience any ordering.
S-type samples also have very large coordination numbers, $8\le z \le 9$. This latter characteristic is a clear indicator of partial
crystalline order, as one necessarily has $z\le 6$ in generically disordered situations. The denser crystal arrangements, face-centered cubic (fcc)
and hexagonal compact (hcp) (and stacking variants thereof), have $z=12$. For D simulations, anomalous values of $\Phi$ and $\sigma_{12}$ appear after strains of order $5$.

In order to detect crystalline order more quantitatively, we use the standard order parameters $Q_6$ and $Q_4$ employed in~\cite{VCKB02,ASS05,LGROT02,iviso1}. Values of the pair $(Q_4,Q_6)$ can be used to distinguish different local environments. In~\cite{iviso1}, following~\cite{ASS05}, the frequency of occurrence of ranges of values  $(0.191 \pm 0.05,  0.574 \pm 0.05)$ and $( 0.097 \pm 0.05,  0.485 \pm 0.05)$, respectively corresponding to fcc-like and hcp-like configurations around one grain, were recorded. In the present case, most samples had very similar proportions of hcp-like and fcc-like local arrangements as in the RCP states studied in~\cite{iviso1}: about 12\% of beads fall
in the hcp category, and fcc-like ones are virtually absent. The exceptions are the samples with anomalous, crystal-like properties, for which,
while none of the beads has an fcc-like environment in that sense, the proportion of the hcp-like category raises to about 60\% in S samples
and to 40\% in D ones.

\begin{figure}
\centering
\includegraphics[width=6cm]{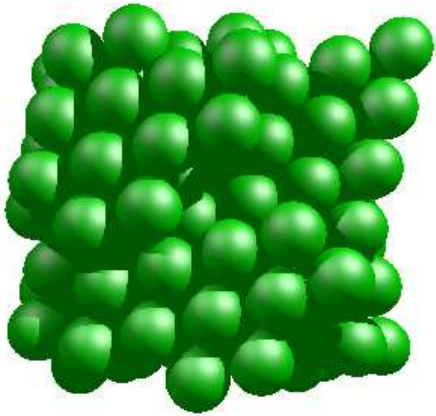}
\caption{\label{fig:crystal}
Crystalline order induced by shear in one S sample with $N=256$.
}
\end{figure}

A direct visualization, Fig.~\ref{fig:crystal}, reveals strikingly ordered configurations.
A tentative conclusion to those preliminary observations is that the small samples tend to crystallize on somewhat shear-distorted hcp lattices. One convenient characterization of order that is not sensitive to the distortion of crystalline patterns was suggested in~\cite{VCKB02}, and used in~\cite{iviso1}. With this method, more than 90\% of the particles of the anomalous samples are declared to belong to crystalline regions.


\end{document}